\RequirePackage{fix-cm}
\documentclass[12pt]{article}
\usepackage[fontsize=13pt]{scrextend}
\usepackage{setspace,graphicx,epstopdf,amsmath,amsfonts,amssymb,amsthm, lscape, flafter, hyperref,float, caption, tabularx, marginnote,datetime,enumitem,subfigure,rotating,fancyvrb, mathtools, booktabs, changepage, setspace, placeins, threeparttable, ragged2e, stmaryrd, array}
\usepackage{amsmath}
\usepackage[longnamesfirst]{natbib}
\usepackage[export]{adjustbox}
\newcolumntype{Y}{>{\raggedleft\arraybackslash}X}% raggedleft column X
\usdate
\usepackage[a4paper,bindingoffset=0.2in,%
            left=0.8in,right=0.8in,top=1in,bottom=0.8in,%
            footskip=.35in]{geometry}
            
\newcommand{\owntag}[1]{\stepcounter{equation}\tag{\theequation, #1}}
\makeatother

\usepackage{indentfirst} % Indent first sentence of a new section.
\usepackage{endnotes}    % Use endnotes instead of footnotes

\begin{document}

\setlist{noitemsep}  % Reduce space between list items (itemize, enumerate, etc.)
\onehalfspacing      % Use 1.5 spacing
% Use endnotes instead of footnotes - redefine \footnote command
\renewcommand{\footnote}{\endnote}  % Endnotes instead of footnotes

\author{\large{Mykola Pinchuk}\thanks{\rm Simon Business School, University of Rochester. Email: Mykola.Pinchuk@ur.rochester.edu. \newline I would like to thank Shuaiyu Chen, Yixin Chen, Yukun Liu, Alan Moreira, Robert Novy-Marx, Christian Opp, Xuyuanda Qi, Jerold Warner, Yuchi Yao and Yushan Zhuang for helpful comments. All errors are my own.}}

\title{\bf Monetary Uncertainty as a Determinant of the Response of Stock Market to Macroeconomic News}

\date{21 November 2022}  

\maketitle
\thispagestyle{empty}

\bigskip

\normalsize

\vspace{1cm}

\centerline{\bf Abstract}

\vspace{0.5cm}

\begin{onehalfspace}  % Double-space the abstract and don't indent it
  \noindent This paper examines the effect of macroeconomic news announcements (MNA) on the stock market. Stocks exhibit a strong positive response to major MNA: 1 standard deviation of MNA surprise causes 11-25 bps higher returns. This response is highly time-varying and is weaker during periods of high monetary uncertainty. I decompose this response into cash flow and risk-free rate channels. 1 standard deviation of good MNA surprise leads to plus 30 bps returns from the cash flow channel and minus 23 bps per 1\% of monetary uncertainty from the risk-free rate channel. The risk-free rate channel is time-varying and is stronger when monetary uncertainty is high. High levels of monetary uncertainty mask the strong positive response of stocks to MNA, which explains why past research failed to detect this relation.  
\end{onehalfspace}
\medskip

\clearpage
%\doublespacing
\setstretch{1.5}

\section{Introduction} \label{sec:Model}

A price of an asset is a function of the information set of investors. Any movement in price must be driven by the arrival of new information. One example of such information is news about the state of the economy, i.e., macroeconomic news announcements (MNA). Counterintuitively, an extensive body of research over the recent decades failed to find a large directional response of stock prices to MNA surprises. The lack of this evidence is puzzling since it raises questions about the relevance of macroeconomic conditions to stock prices and contradicts the intuition of practitioners and academics.
\paragraph{}
The aim of this paper is to establish a set of stylized facts on the response of financial assets to MNA and to use this evidence to refine our thinking about asset prices. Utilizing crisper identification and high-frequency data allows me to detect patterns of price response to MNA, which eluded researchers for decades. My results suggest a large positive response of both stocks and risk-free rates to MNA surprise and reaffirm the strong link between macroeconomic conditions and prices of financial assets. 
\paragraph{}
First, I document that several types of MNA have a large effect on both volatility and the direction of aggregate stock returns. In a tight window around these announcements, 9-20\% of movement in stock prices is explained by MNA surprises. Response to these MNA alone is as large as 35\%-70\% of price variation during periods without MNA. Stocks exhibit a strong positive response to major MNA: 1 standard deviation of good surprise causes 11-25 bps higher returns. Unsurprisingly, these are MNA, closely followed by the financial press: non-farm payroll, PMI, retail sales, and consumer confidence.
\paragraph{}
Second, I report that the response of stocks to MNA crucially depends on the level of uncertainty about monetary policy. This allows me to reinterpret puzzling findings that stocks tend to respond somewhat negatively to positive growth news (Boyd, Hu, Jagannathan 2005). Good growth news implies both higher expected dividends and a higher future risk-free rate. Therefore, these two channels affect stock prices in opposite directions.  When uncertainty about monetary policy is low, the risk-free rate is not expected to vary a lot, making the risk-free rate less sensitive to changes in the economic environment. Thus we can use periods of very low monetary uncertainty to cleanly measure the cash flow channel of stock response to macroeconomic news. 
\paragraph{}
I show that the positive response of stocks to good (i.e., higher growth) MNA surprise is much stronger when monetary uncertainty is low. A high level of monetary uncertainty makes the positive response of stocks to MNA weaker. At a high enough level of monetary uncertainty, the relation between stock returns and MNA surprise may even reverse due to the larger strength of the monetary channel. Very high monetary uncertainty during the 1970s-1980s is likely to be the reason previous research failed to find a positive response of stocks to MNA surprise. I extend my findings to the 1980s and show that stock response likely was negative due to a very large negative risk-free rate channel. A simple theoretical framework allows me to quantitatively assess the contribution of these two channels to the reaction of aggregate stock prices to most types of MNA. After controlling for the monetary channel, 1 standard deviation surprise in major MNA causes 30 bps higher returns, i.e., price movement, equal to 90\% of the standard deviation of returns during similar 30-minute periods on days without MNA. 
\paragraph{}
Third, I document that none of the major MNA surprises is priced in the cross-section of expected stock returns. I use MNA surprise as a proxy for macroeconomic shock, associated with each MNA type and estimate the sensitivities of individual stocks to these shocks. None of the 4 major MNA shocks produces a significant return spread. This finding is consistent with the idea that risk premium is not significantly affected by MNA, implying that most changes in discount rate come from the risk-free rate. The whole yield curve appears to shift up or down in response to the MNA surprise. The effect is larger for short maturities and slowly declines at the long end of the term structure.
\paragraph{}
The literature on the behavior of stock prices during periods of MNA dates back at least to Schwert (1981). Pearce and Roley (1985) find that money supply is the only MNA, which affects stock returns. Cutler, Poterba, and Summers (1989) use vector autoregression on monthly data to quantify the fraction of returns, driven by macroeconomic news. McQueen and Roley (1993) suggest that response of the stock market to MNA may depend on the business cycle. They find a positive response to good news during periods of low growth and a negative response during a boom. Flannery and Protopapadakis (2002) argue that the nonlinearity and time dependence of the effects of MNA makes them hard to detect. They focus on the response of market volatility to MNA and find that money supply and inflation are the most important MNA. Boyd, Hu, and Jagannathan (2005) study market response to unemployment news and find that on average bad unemployment news has a positive effect on stocks. They explain this result by the hypothesis that during expansions market perceives unemployment news as news about interest rates rather than economic growth. Law, Song, and Yaron (2020) use the New Keynesian model to argue that the stock response to MNA surprise is the strongest during periods of a low output gap. My paper reinterprets the results in the papers above by showing that monetary uncertainty is the driver of state dependence of stock response to MNA. Since monetary uncertainty is often the lowest during early recovery, previous research misattributed time variation in this response to business cycles rather than monetary uncertainty. I show that the stage of the business cycle has little explanatory power after controlling for monetary uncertainty.
\paragraph{}
By uncovering and quantifying the large monetary channel of the stock response to major MNA, this paper contributes to monetary economics. The last two decades saw increased interest in asset pricing effects of monetary policy. Bernanke and Kuttner (2005) used high-frequency identification to argue that stocks react negatively to Fed Funds rate surprises. There is a growing body of research on the implications of zero lower bound (ZLB) for monetary transmission and asset prices. Swanson and Williams (2013, 2014) and Zhou (2014) focus on the response of forward rates to monetary news at ZLB. My results suggest that ZLB is an extreme case of low monetary uncertainty and is useful to isolate out cash flow channel of MNA response.
\paragraph{}
The paper is organized as follows. Section 2 explains the advantages of event studies for identifying the relation between asset prices and macroeconomic conditions. Section 3 contains a simple theoretical framework, which allows us to decompose the response into channels. Then Section 3 explains the intuition behind the idea of monetary uncertainty and its time variation. Section 4 discusses data sources, identifies types of MNA, and discusses the construction of monetary uncertainty proxy. Section 5 focuses on the response of stocks to major MNA and decomposes this response into cash flow and monetary channels using a simple statistical framework. Section 6 explores the cross-section of stocks by their response to major MNA. Section 7 concludes.

\section{Estimation of the response using event study}

Asset pricing has a long tradition of estimating exposures of financial assets to macroeconomic variables. Most studies use low-frequency regressions to perform such an estimation. For example, to estimate the exposure of stock to consumption growth, we usually regress its returns on contemporaneous consumption growth at monthly (quarterly) frequency. While the simplicity of this approach makes it very appealing, it suffers from some drawbacks. Examples of such limitations are a lack of causal interpretation of estimates, low statistical power, and the inability to distinguish between similar macroeconomic variables. Another problem with this approach is its inability to account for market expectations in a reasonably precise way. As emphasized by Brunnermeier et al. (2021), since returns of financial assets should be driven by changes in expectations, controlling for the expectations is critically important. The traditional approach usually uses either first differences or residuals from some time-series model, such as VAR, to measure innovations in macroeconomic variables. Multiple studies (Ang et al., 2007, Rossi and Sekhposyan, 2015) suggest that investors` surveys provide a better proxy for expectations, compared to time series models. To the best of my knowledge, there is no data on investors' forecasts, made during the last days of the month t-1 for macroeconomic variables in month t. Thus controlling for investors` expectations is not feasible within the traditional framework.
\paragraph{}
High-frequency event studies provide a somewhat overlooked alternative to the traditional approach for estimating exposures of financial assets to macroeconomic variables. Event studies in finance date back at least to Fama, Fisher, Jensen, and Roll (1969) and Ball and Brown (1968). As intraday data and investors` surveys became available during the last 25 years, we can use event study methodology to estimate responses of asset prices to macroeconomic variables in a way, which mitigates most of the limitations of the traditional approach. Such high-frequency event study involves regressing asset returns during a short intraday window, centered on news announcement, on announcement surprise. An announcement surprise is a difference between the announced value of the macroeconomic variable and a mean value from the investors` survey. 
\paragraph{}
Such high-frequency event studies have several advantages over traditional low-frequency analysis:
\begin{enumerate}
    \item {Higher signal-to-noise ratio.}
    \item {Controlling for market expectations.}
    \item {Ability to distinguish between the effects of highly correlated macroeconomic variables.}
    \item {Higher statistical power from combining multiple MNA of the same type. }
    \item {Causal interpretation of results.}
\end{enumerate}
\paragraph{}
First, measuring the response of asset prices to MNA over a short intraday window allows us to isolate the shock, associated with MNA surprise. This leads to higher statistical power and enables us to detect asset price responses, previously obscured by noise in asset returns. 
\paragraph{}
Second, Brunnermeier et al. (2021) emphasize the importance of accounting for market expectations when analyzing asset price movements. Since multiple entities (Bloomberg, Reuters, etc) conduct investors` surveys immediately before the release of MNA, it is possible to measure the response of asset prices to actual changes in market expectations.
\paragraph{}
Third, most macroeconomic variables are measured at a monthly or quarterly frequency. Since many assets do not have long time series of their returns, this often results in a small sample size. Hence estimating the exposure of financial assets to macroeconomic variables often suffers from low statistical power. Combining multiple MNAs of a similar type (e.g., economic growth) allows to effectively increase the sample size by a factor of 2-10 and achieves higher statistical power. 
\paragraph{}
Fourth, one of the problems with low-frequency analysis, commonly used to estimate loadings of an asset on macroeconomic variables, is the high correlation between most macroeconomic variables. This makes it difficult to distinguish between the effects of correlated macroeconomic variables. Since most MNA releases do not coincide in time, an event study can easily identify the most relevant MNA. By definition, irrelevant MNA does not induce any market reaction upon release.
\paragraph{}
Finally, estimates of the exposure of financial assets to macroeconomic variables using traditional analysis are just correlations. Since at low frequency many macroeconomic variables may be affected by financial market movements, such estimates suffer from reverse causality. Furthermore, as explained above, these estimates may be hard to attribute to a specific macroeconomic variable. Using high-frequency event studies allows us to mitigate this endogeneity and gives causal interpretation to such estimates. If the stock index jumps during the minute when we receive good employment news, the such jump must be caused by MNA surprise.

\section{Theoretical Framework}

\subsection{Asset response to a growth shock}

This section describes a simple framework, which allows me to decompose the response of stocks to MNA shocks into cash flow channel and monetary channel.
% \paragraph{}
% Consider a problem, in which the market tries to learn the value of unobserved short-term interest rate $\hat{i}_t$. To do so, it relies on its prior knowledge as well as on the signal $m_t$, contained in MNA. Thus, the problem is to dynamically use noisy signal $m_t$ to extract useful information about unobserved state $\hat{i}_t$. 
\paragraph{}
Consider the response of the aggregate stock market to pure growth shock $\epsilon_t$. In my empirical setting, such shock corresponds to MNA surprise. According to the one-period pricing identity (1), the stock price is a function of expected dividends, risk-free rate, and risk premium.
\begin{equation}
    P_t = \frac{\mathbb{E}_t[D_{t+1}]}{i_t + RP_t}.
\end{equation}
Thus any price movement must come from one or more of these three channels. Most MNAs in my sample have a natural interpretation as shocks to expected growth, i.e., cash flows. However, due to the dual mandate of the Fed, it is supposed to use monetary policy to offset cyclical fluctuations. Therefore, any changes in macroeconomic expectations are likely to affect future monetary policy. Hence each MNA affects stocks through at least 2 channels: the growth channel and the monetary channel. Furthermore, if MNA is related to risk premium, it could move prices through all 3 channels. 
\paragraph{}
Consider growth news $\epsilon_t$, which does not affect risk premium. Then, according to log-linearized (1), it should move the price through expected cash flow and risk-free rate ($\Delta i_t$) channels:
\begin{equation}
    R_t = a_1*\epsilon_t + a_2*\Delta i_t + e_{1,t}.
\end{equation}
The risk-free rate channel reflects changes in expected monetary policy due to growth news $\epsilon_t$. Kalman Filter described in Appendix B suggests such a response should depend on monetary uncertainty $MU_{t-1}$:
\begin{equation}
    \Delta i_t = (\gamma_0 + \gamma_1*MU_{t-1})*\epsilon_t + e_{2,t}.
\end{equation}
Plugging (3) into (2) and rearranging the terms yields (4):
\begin{equation}
    R_t = [a_1+a_2*\gamma_0]*\epsilon_t + [a_2\gamma_1]*MU_{t-1}*\epsilon_t + e_{3,t}.
\end{equation}
\paragraph{}
Equation (B.10) from Appendix B suggests that $\gamma_0=0$. In subsequent sections, I empirically verify that $\gamma_0=0$. This allows me to simplify (4):
\begin{equation}
    R_t = [a_1]*\epsilon_t + [a_2\gamma_1]*MU_{t-1}*\epsilon_t + e_{3,t}.
\end{equation}
The equation above provides an intuitive decomposition of stock response into two components. $a_1$ shows the strength of the response to growth shock $\epsilon_t$. $a_2\gamma_1$ shows the sensitivity of stock response to the change in short-term interest rates. In some sense, this decomposition relies on using monetary uncertainty $MU_{t-1}$ to instrument for the exogenous component of the risk-free rate. I can estimate equation (5) by regressing stock returns on MNA surprise $\epsilon_t$, monetary uncertainty $MU_{t-1}$, and their interaction term. Before running this regression, I will estimate (3) and verify that $\gamma_0=0$.
% in the empirical section, mention that I estimate (13) and get gamma_0 = 0. then mention that a_2gamma_1 is the sensitivity of stocks to short-term rate and the coefficient is consistent with bk05 and related papers. so it passes the sanity check.
\paragraph{}
This decomposition relies on two assumptions. First, I assume that MNA surprise does not affect the risk premium. Section 6 shows that this assumption is likely to hold for all 4 major MNA in the cross-section. Second, I assume that the cash flow channel of the response of stocks to growth shock $\epsilon_t$ does not depend on the level of monetary uncertainty. Results in Chapter 5 show that if such dependence exists, it goes in the opposite direction to the monetary channel. So controlling for such dependence can make the estimates of the two channels even larger.
\paragraph{}
Intuitively, this framework is based on time variation in the ability of the Fed to use monetary policy tools to affect the economy. Low monetary uncertainty periods indicate that the Fed has the limited ability or willingness to use monetary policy tools. Thus we should not expect a large movement in interest rates in response to changes in the macroeconomic environment. In the limiting case of zero monetary uncertainty, any incoming macroeconomic news $\epsilon_t$ is interpreted as pure cash flow news. By proxying for the strength of the risk-free rate channel, monetary uncertainty allows us to distinguish between cash flow and risk-free rate channels.

\subsection{Monetary Uncertainty} \label{sec:Model}

The Fed determines monetary policy according to its mandate, trying to achieve price stability and maximum employment. Thus as the economic environment changes, the Fed is supposed to adjust its monetary policy. In practice, it usually means cutting the Fed Funds rate during recessions and raising it during economic booms. Naturally, the response of the Fed to changes in economic conditions usually has some degree of uncertainty. Such monetary uncertainty can vary due to two reasons. First, the Fed is often constrained in its ability to use its monetary policy tools to shape economic outcomes leading to lower monetary uncertainty. Second, the Fed occasionally changes the manner in which it implements monetary policy. For example, attempts to increase transparency and continuity of monetary policy often lead to a decrease in Fed flexibility and lower monetary uncertainty. I explain these two reasons in three paragraphs below.
\paragraph{}
The Fed's ability to affect the economy via monetary policy varies over time. The main tool of the monetary policy is the Fed Funds rate. The Fed stimulates the economy by slashing the Fed Funds rate and cools the economy by raising the Fed Funds rate. Thus a low level of the Fed Funds rate limits the ability of the Fed to affect the economy. At zero lower bound (ZLB) Fed Funds rate is zero and the Fed has a very limited ability to conduct monetary policy by changing the policy rate. ZLB implies very low monetary uncertainty since the Fed Funds rate is unlikely to move significantly in either direction. While theoretically possible, a negative Fed Funds rate is very unlikely due to implementation difficulties. Furthermore, at ZLB an increase in the Fed Funds rate at any given point in time is less likely as well. From the standpoint of the baseline New Keynesian model, a zero Fed Funds rate means that the Fed is likely to have a negative desired Fed Funds rate. Thus positive growth news of a small magnitude is likely to lead to the Fed updating its desired Fed Funds rate to a less negative number and leaving the actual Fed Funds rate at zero. Thus a low interest rate implies low monetary uncertainty. 
\paragraph{}
A level of Fed Funds rate may not be the only constraint on Fed's ability to adjust monetary policy. A higher level of public debt relative to GDP could prevent the Fed from raising the policy rate for the fear of causing an unsustainably high cost of serving public debt. Between 1980 and 2019 debt to GDP ratio rose from 31\% to 104\%, possibly contributing to lower monetary uncertainty. 
\paragraph{}
In conducting monetary policy, the Fed faces tension between two objectives. On the one hand, the Fed wants to be able to adjust monetary policy rapidly as economic conditions change and as its understanding of the state of the economy evolves. On the other hand, The Fed tries to maintain a consistent and predictable monetary policy to minimize the risk of financial turmoil from sudden policy changes. Different Fed Chairmen chose a different balance between these two objectives. The Fed during the periods of Volcker and Greenspan (1979-2006) maintained minimal communication policy in order to maximize Fed's flexibility. Starting from Bernanke, Fed Chairmen put more emphasis on detailed communication of monetary policy decisions and policy projections. For example, in 2011 in an attempt to increase transparency and clarity of monetary policy, the Fed introduced press conferences following FOMC meetings. The ability of financial media to publicly question the Fed Chairman on the monetary policy details and projections is likely to decrease monetary uncertainty. Moreover, such communication often involves forward guidance, often viewed as a soft commitment to the short-term path of interest rates, further lowering monetary uncertainty. Clarity and continuity of monetary policy decisions are other reasons for time-varying monetary uncertainty. For example, over 1979-1981 Fed Funds rate fluctuated widely, ranging from 8\% to 22\%. Perceived unpredictability of the Fed under Volcker's leadership would likely imply high monetary uncertainty stemming from Fed decision-making. More recently, a lack of consistency in the monetary policy over 2020-2022 as well as policy errors contributed to a high monetary uncertainty as of the time of writing of this draft. Monetary uncertainty is usually high when past policy errors necessitate the rapid readjustment of monetary policy.

\section{Data and News Types} \label{sec:Model}

I obtain MNA data from Bloomberg. Over 1997-2019 I download all MNA for variables, which are collected at a monthly frequency and are primarily related to economic growth. There are 17 such announcement types. Table 1 describes the types and exact times when these MNAs are released. In a few cases macroeconomic variables were not released for some months, so there is a small variation in the number of announcements for each MNA. For most MNA, my sample includes 272-278 news releases.
\paragraph{}
Since asset prices reflect beliefs about the future, to capture their forward-looking nature, I need some proxy for expectations. I use Bloomberg survey data to proxy for expected values of these macroeconomic variables before the MNA release. Bloomberg collects these data for 2 weeks before a news release, with most forecasts made within 5 days before an announcement. Bloomberg stops updating forecasts 2 hours before the announcement. This allows me to obtain a timely proxy for a news surprise. I compute news surprise as a difference between the announced variable and its survey median.
\paragraph{}
Table 1 shows that many MNA are released at 8:30 am when stock exchanges are closed. Therefore, to estimate the reaction of financial assets to such announcements, I use futures, which are traded around the clock. To proxy for the aggregate stock market, I use E-Mini S\&P 500 futures. To gauge the response of the risk-free rate, I use Fed Funds futures as well as Treasury rates at maturities of 2, 5, and 10 years. My sample of Fed Fund futures and Treasury rates starts in 1997. For additional results, I use daily prices of Treasury securities at all possible maturities.
\paragraph{}
I use S\&P 500 futures, Fed Funds futures, and Treasury rates data at a 1-minute frequency. As I show in Section 5, using high-frequency data is critically important to identify the response of asset prices to MNA. In addition to aggregate data, I download prices of constituents of S\&P500 from TAQ. These data span 1997-2019 and are aggregated at a 5-minute frequency. One potential drawback of using TAQ is that trade data starts at 9:30 am, making it somewhat difficult to estimate stock response to early-morning MNA.
\paragraph{}
I use Bloomberg data on Black implied volatilities of US LIBOR rate swaptions to proxy for implied volatilities of Treasury rates. Since implied volatility of swaptions describes possible fluctuation per 1\% of interest rate, I multiply this variable by the level of corresponding interest rate. For example, to proxy for the implied volatility of the 2-year Treasury rate, I use Black implied volatility of swaptions with a 1-year maturity and 2-year tenor. In other words, I use options on interest rate swaps with a length of two years and an expiration of 1 year. Then I multiply this implied volatility by the level of the 2-year Treasury rate. 
\paragraph{}
This paper uses the implied volatility of the 2-year Treasury as a proxy for monetary uncertainty. Figure 1 describes the time variation of monetary uncertainty. To extend my results to the period between 1980 and 1997, I use realized volatility of 2-year Treasury notes.
\paragraph{}
The idea that implied volatility of 2-year Treasury note yields captures monetary uncertainty relies on the notion that 2-year yield reflects monetary policy. Since the recent financial crisis, literature in both monetary economics and finance has usually used the 2-year Treasury yield as a proxy for monetary policy. Hanson and Stein (2015) use a 2-year rate to "capture full expected path of fed funds rate over the coming quarters in a simple and transparent manner". Nakamura and Steinsson (2018) use a 2-year nominal yield as a policy indicator while estimating the effect of monetary policy shock. Before 2009 most papers (Rigobon and Sack 2004, Bernanke and Kuttner 2005, Gurkaynak, Sack and Swanson 2005) proxied for monetary policy shock using Fed Funds futures and Eurodollar futures with expiration from 3 months to 1 year. Due to a prevalence of zero lower bound after 2009, it makes sense to use longer maturity to capture the forward guidance component of monetary policy.

\section{Stock Market Response to MNA} \label{sec:Model}

\subsection{Main Results} \label{sec:Model}

\paragraph{}
Ernst et al. (2020) show that any study which explores a large set of MNAs could suffer from a data mining problem by selecting a few MNAs. To avoid multiple hypothesis testing concerns, I first identify a set of MNAs, which have a significant effect on the market, and then study only those MNAs. 
\paragraph{}
To be relevant for financial markets, MNA must lead to increased volatility immediately after the announcement. To measure such an increase, I look at the absolute value of market returns during the 30-minute window around news releases. Unless stated otherwise, all analysis in this paper uses the window, starting 10 minutes before the announcement and ending 20 minutes after it. Then I compare average absolute returns within this window to average absolute returns during the same time of day over days without MNA. Relevant MNA should produce a significantly higher absolute value of returns compared to the no-news baseline.  
\paragraph{}
Table 2 describes this volatility. To test whether MNA generates increased variation in returns, I run both the T-test and Mann-Whitney-Wilcoxon tests between MNA and no-news periods. To be considered relevant, MNA must have a p-value below 0.05 in the T-test. According to this metric, there are 5 relevant MNAs: 
\begin{itemize}
    \item {Non-farm payroll (NFP).}
    \item {ISM Manufacturing (PMI).}
    \item {Retail Sales.}
    \item {Construction Spending.}
    \item {Conference Board Consumer Confidence.}
\end{itemize}
After identifying 5 types of MNA to explore, I report their summary statistics in Table 3. It documents the distribution of aggregate stock returns during MNA, news surprises as well as levels of several macroeconomic variables, which will be used in subsequent analysis. For each MNA, I normalize its news surprise by its full-sample standard deviation. All results are robust to using a rolling or expanding window to estimate standard deviation for scaling.
\paragraph{}
Then I explore the reaction of the market to these MNA. Since NFP and unemployment news are announced as a part of a single release, Table 4 contains 6 types of MNA. To assess this response, I run a simple OLS:
\begin{equation}
    R_t = a + b*Surprise_t + \epsilon_t.
\end{equation}
Table 4 reports coefficient b and adjusted $R^2$ for each type of MNA. There are 4 types of MNA, which have a large directional effect on the stocks: NFP, PMI, Retail Sales, and Consumer Confidence. Their coefficient estimates are highly statistically significant with t-statistics above 4. More importantly, economic magnitudes are large: 1 standard deviation of good news leads to 11-25 bps higher returns. $R^2$ from these regressions imply that these 5 MNA surprises can explain 9-20\% of return variation during announcement windows.
\paragraph{}
Another way to see the large economic importance of these responses is to compare the magnitude of returns, caused by news surprise, to average price movement during similar periods without MNA. 1 standard deviation of S\&P 500 returns over the no-news window at 8:30 (10:00) is 34 (37)bps. In other words, 1 standard deviation of MNA news surprise creates price movement, equal to 35-75\% of price movement on comparable periods without MNA. 
\paragraph{}
The direction of stock response is another interesting result. While, intuitively, good (i.e., higher growth) macroeconomic news should lead to higher stock prices, past research (McQueen and Roley 1993, Flannery and Protopapadakis 2002, Boyd, Hu, and Jagannathan 2005) failed to detect such a relation. These papers find either no significant unconditional relation between stock returns and news surprises or even negative relation. Thus they focus on more complex relations such as the effect of the news on market volatility or the differential effect of MNA surprises over a business cycle. For example, Boyd, Hu, and Jagannathan (2005) argue that, on average, good unemployment news is bad news for stocks, since such news is seen as bad news during expansions due to a larger effect on interest rates than expected growth. But even when considering more complex relations, these papers struggle to find patterns with high statistical and economic significance.
\paragraph{}
My identification approach has 2 main advantages over these papers, which allows me to estimate the response more accurately.
\paragraph{}
First, using intraday data dramatically improves the signal-to-noise ratio. Given the high liquidity of the stock market over the recent decades, it is reasonable to expect that most of the important and widely followed news will be incorporated into prices within minutes. This is especially true for macroeconomic news due to its standardized and repetitive nature. Using a 30-minutes or 1-hour window to measure the effect of monetary policy changes on financial markets has been standard for more than 15 years (Gurkaynak, Sack, and Swanson 2005). For some reason, this approach is not yet dominant in the analysis of the connection between asset prices and macroeconomic variables.
\paragraph{}
Table 5 illustrates the benefits of using a 30-minutes window to identify the effect of MNA on the example of PMI news. While moving from daily frequency to a 30-minute window (i.e., column 1 versus column 4) has little effect on the coefficient estimate, it dramatically reduces standard error. Using a daily window, we would find a barely significant positive response with $R^2$ of 1\%. So it would appear that this pattern is too small to be worth studying. But using a 30-minutes window allows us to uncover a highly significant response with a t-statistic of 5 and $R^2$ of 9\%.
\paragraph{}
The related question is the width of the window to measure this response. My paper uses a 30-minute window, which is common in event study literature. To understand the effect of varying window width on the response, I plot an evolution of stock prices around announcements in Figure 2. Around 90\% of the response happens during the first minute. After the first 5 minutes upon the announcement, the price seems to stay at a post-announcement level. The only possible exception from this result is the response of stocks to the most positive MNA surprises. Figure 2 shows evidence of overreaction to such surprises: prices seem to drift back more than halfway toward their pre-announcement level. I observe this pattern only for the most positive news surprises. Overall, this analysis suggests that my results are not sensitive to the choice of window width. Using a shorter window (e.g., 5 minutes around MNA) could help with statistical power, but is uncommon in the literature. Using a longer window (e..g, 2 hours) may alleviate some concerns about reversal, but will decrease the signal-to-noise ratio.

\paragraph{}
The second methodological improvement is in using surveys of market participants to proxy for their expectations. Most papers, studying MNA, use either time-series models or low-frequency surveys to construct expectations, needed to compute news surprises. I use Bloomberg surveys, which are continuously updated up to 2 hours before MNA. A better proxy for market expectations leads to a more accurate measurement of a news surprise.
\paragraph{}
Results in Table 4 suggest that unemployment and construction news do not have a highly significant effect on aggregate stock returns. This evidence suggests that the market views NFP as a more important indicator of the state of the labor market than unemployment. This result is consistent with the Congress Testimony of Alan Greenspan on Feb 11, 2004, when he suggested investors pay closer attention to NFP than unemployment. Unlike unemployment, NFP accounts for both the labor force participation rate and the number of job seekers. Thus focus on unemployment as opposed to NFP may be another reason why past research failed to uncover a strong response of stocks to labor market news.
\paragraph{}
In the remainder of the paper, I restrict my analysis to 3 types of MNA, which induce large market response: NFP, PMI, and Retail. Not surprisingly, these types of news are among the most closely-followed news by the financial press. 
\paragraph{}
After documenting the large response of stocks to MNA surprises, I explore its dependence on the economic environment. McQueen and Roley (1993), Flannery and Protopapadakis (2002), Boyd, Hu, and Jagannathan (2005), and Law, Song, and Yaron (2021) suggest that this response may vary across stages of the business cycle. To test this hypothesis, I divide the sample into 2 parts by the rates of economic growth, proxied by the Chicago Fed National Activity Indicator (CFNAI). Additionally, I look at this response across samples with low and high interest rates.
\paragraph{}
Table 6 contains the results for the 4 most important types of MNA (NFP, PMI, Retail, and Consumer Confidence). This table reports estimates from regression (6) in subsamples, divided by growth and interest rates. For every MNA, we observe the following pattern. Unconditionally, news surprises have a positive effect on stocks. This effect appears stronger during recessions, though this difference is not very large (6 bps). 
\paragraph{}
The last two columns in Table 6 show that stock response to MNA surprise differs dramatically across periods of low monetary uncertainty versus high monetary uncertainty. While positive growth news is still good news for the market, the response is much weaker during periods of high monetary uncertainty. The magnitude of this response is almost 3 times stronger during the period of low monetary uncertainty. In other words, when the short-term interest rate is not expected to change, good macroeconomic news is interpreted very optimistically by the market. On the other hand, when the monetary policy outlook is uncertain, good growth news has muted effect on the market.
\paragraph{}
Whereas these results are different from McQueen and Roley (1993), Flannery and Protopapadakis (2002), and Boyd, Hu, and Jagannathan (2005), it is not hard to reconcile them. The dependence of stock response on the stage of the business cycle appears to be a manifestation of the monetary uncertainty effect. Post-recession recovery coincides with periods of low rates and low monetary uncertainty. Thus previous findings of a strong positive response during recession and recovery were likely driven by a low interest rates uncertainty. Failure of these papers to find a positive response of stocks to MNA surprise in their full sample was mostly due to very high monetary uncertainty in the 1980s. My sample, starting in 1997, does not contain periods of monetary uncertainty, high enough to flip the sign of the response.
\paragraph{}
Intuitively, MNA should affect stocks through 3 channels: expected cash flows, risk-free rate, and risk premium. A good growth shock will raise expectations of dividends but is likely to increase the discount rate through its risk-free component. The next subsection discusses the decomposition of these effects using the theoretical framework from Section 3.

\subsection{Decomposition of the response} \label{sec:Model}

I decompose stock response to MNA surprise into channels using (7), an empirical counterpart to (5). Table 7 reports the estimates of the coefficients in the regression with an interaction term between MNA surprise and monetary uncertainty (MU):
\begin{equation}
    R_t = a + b*Surprise_t + c*\text{MU}_{t-1} + d*Surprise_t*\text{MU}_{t-1} + \epsilon_t.
\end{equation}
I proxy for monetary uncertainty using implied volatility of a 2-year Treasury rate 1 day before MNA. Controlling for interaction between monetary uncertainty and MNA surprise doubles the coefficient on MNA surprise. The coefficient on the interaction term is negative (-23 bps) and highly significant (t-statistic above 4). Consistent with the results in Table 6, adding the interaction term of surprise with a proxy for the stage of the business cycle (Chicago Fed National Activity Index, CFNAI) does not affect results.
\paragraph{}
One standard deviation of news surprise leads to 30 bps positive returns when the level of monetary uncertainty is zero. As monetary uncertainty increases, this positive response becomes weaker. When the implied volatility of the Treasury rate is equal to 1.3\%, the interaction term perfectly offsets the positive channel, and surprise does not affect stock prices. At levels of the implied volatility of the Treasury rate above 1.3\%, we are likely to see a negative response: stocks will fall in response to good growth news. 
\paragraph{}
These results imply that the interaction term proxies for the strength of the monetary channel, while the coefficient on MNA surprise captures the cash-flow channel. Intuitively, when monetary uncertainty is very low (e.g., at ZLB), the Fed does not respond to changes in economic conditions. Zero monetary uncertainty allows us to shut down the risk-free rate channel and estimate the pure cash-flow channel. Thus it is natural to expect that monetary uncertainty could instrument for the strength of the risk-free rate channel.
\paragraph{}
The last two columns in Table 7 show the response of the 2-year Treasury yield to the MNA surprise. Following the literature (Hanson and Stein 2015), I use front-month 2-year Treasury rate futures to proxy for changes in the expected path of monetary policy. Results in column 4 mean that the risk-free rate increases in response to good news. Risk-free rate changes by 2.2 bps in response to 1 standard deviation MNA surprise. 5-year and 10-year Treasury rates exhibit a similarly strong response to MNA surprise (1.7-2.3 bps), suggesting that positive MNA surprise shifts the whole yield curve up. The positive interaction term in the last column implies that the shoer-term interest rate is more sensitive to MNA surprise when monetary uncertainty is farther from zero. Overall, these results are consistent with the hypothesis that the monetary channel is stronger when the monetary uncertainty is higher. 
\paragraph{}
Column 5 of Table 7 implies that $\hat{\gamma_0}=0$, so the estimates of coefficients on the MNA surprise in column 2 are $a_1$ and $a_2*\gamma_1$, i.e., cash flow channel and risk-free rate channel respectively. 1 standard deviation of positive MNA surprise raises stock prices by 30 bps due to higher expected cash flows. At the same time, 1 standard deviation of good MNA surprise decreases stock prices by 23 bps per each 1\% of monetary uncertainty $MU_{t-1}$. This is an estimate of the strength of the monetary channel.
\paragraph{}
The decomposition provides a possible answer to why many papers (McQueen and Roley 1993, Flannery and Protopapadakis 2002, Boyd, Hu, and Jagannathan 2005), failed to uncover the large effect of MNA on the stock market. The direction and magnitude of stocks' response to MNA depend on the relative strength of cash flow and monetary channels. When monetary uncertainty is high, the monetary channel becomes stronger and can offset the cash flow channel. When monetary uncertainty exceeds 1.3\%, the direction of the response may even reverse. Since the papers above use a sample, starting in the 1970s, the monetary uncertainty in their sample is very high and is likely to have been above the level at which the direction of the response reverses. Hence my results can explain the finding of Boyd, Hu, and Jagannathan (2005) that, on average, stocks react negatively to good unemployment news. It is driven by the fact that in most of their sample risk-free rate channel dominates. But when interest rates are constrained by ZLB and monetary uncertainty is low, the cash flow channel becomes a dominant channel, so we observe a large positive response to good news. My results for 4 major MNAs suggest that 1 standard deviation of good surprise leads to 30 bps positive returns from the cash flow channel. When monetary uncertainty is equal to 1.3\%, this good growth shock will additionally lead to negative returns of a similar magnitude (-23*1.3=-30bps) from the monetary channel. Hence, simple OLS will fail to detect any response during such a period, while the total absolute response from these two channels will amount to 60 bps. Due to the presence of a duration effect, the estimates of the two channels are likely to be the lower bounds of these two channels$^1$. 
\footnotetext[1]{My measure of monetary uncertainty MU is increasing in a level of interest rate. A higher interest rate (i.e., higher MU) means a lower duration. In my framework, the closest variable to duration is a coefficient $a_2$. So, mechanically, the duration effect will decrease an absolute value of $a_2$ for a higher level of MU. This effect goes in the opposite direction to my results, implying that my estimates are lower bounds of the magnitude of the two channels.}
\paragraph{}
Another contribution of this framework is its implications for the effects of monetary policy. While ${a}_1$ and $\gamma_1$ are MNA-specific, $a_2$ should be the same for any MNA. I estimate ${a}_2$ equal to -8. This estimate is broadly consistent with the results in Bernanke and Kuttner (2005), as well as related papers, which estimate the sensitivity of the stock market to interest rate news around the monetary announcement. These papers obtain estimates between -5 and -10, consistent with my findings.
\paragraph{}
While I use 2-year Treasury yield to measure news about monetary policy, the strength of the monetary channel implies that it should affect yields at longer maturities. To answer this question, I estimate the sensitivity of interest rates to MNA surprise using Treasuries with maturities from 1 year to 30 years. Due to a lack of intraday data for most of these rates, I use daily data to estimate such sensitivities. Table 8 and Figure 3 describe the results. Consistent with the idea that MNA surprise contains important news about future monetary policy, the strongest response is at the maturities from 2 to 5 years. As maturity increases, the response of interest rates to MNA surprise weakens and falls to 1.26 bps, a 57\% decrease from 2.91 bps response at 3-year maturity. The positive response at a long end of a yield curve may reflect long-term growth expectations rather than monetary policy. 

\subsection{Extrapolation of the findings to earlier sample and other results} \label{sec:Model}

One of the contributions of this paper is its ability to explain surprising past findings and reconcile them with our economic intuition. The previous subsection laid out this explanation qualitatively. The quantify this explanation, ideally, I have to run response regressions for the 1980-1997 sample. Unfortunately, this data is not available publicly. The second-best solution would be to use estimates from 1997-2019 regressions to measure the strength of the two channels. Then I could multiply the strength of the risk-free rate channel by monetary uncertainty, extrapolating the results back to the 1980s. Unfortunately, I am unable to implement this solution either, since I do not have implied volatilities of Treasuries before 1997.
\paragraph{}
Thus I use the third-best approach, in which I employ realized volatility of Treasuries to proxy for monetary uncertainty. First, I estimate regression, similar to (7), but with realized volatility instead of implied volatility. Then I multiply the coefficient on the interaction term by the level of realized volatility. As a result, I have an extrapolation of a constant cash-flow channel and time-varying risk-free rate channels back to 1979.
\paragraph{}
Table 9 reports the results of running regression (7) with realized volatility. The findings are similar to those with implied volatility: all coefficients have the same signs and statistical significance. Controlling for interaction between realized volatility and MNA surprise increases the coefficient on surprise, while the coefficient on the interaction term is significantly negative. 1 standard deviation of good growth shock leads to 20 bps higher returns due to the cash flow channel and -16 bps per 1\% of interest rate volatility due to the risk-free rate channel. 
\paragraph{}
Figure 4 shows the extrapolation of these results to the 1979-1997 sample. As explained above, the cash flow channel is constant at 20bps, while the risk-free rate channel depends on the level of realized volatility of the 2-year Treasury rate. Figure 4 describes the strength of these two channels over time. As expected, the negative risk-free rate channel was very strong in the 1980s. Between 1980 and 1987 it was larger in absolute value than the positive cash flow channel as shown by a red-shaded area. After 1987 with small exceptions positive cash flow channel dominated the negative risk-free rate channel (the green-shaded area) which is consistent with my main results.
\paragraph{}
Monetary uncertainty is likely to be correlated with broader macroeconomic uncertainty. Periods of higher macroeconomic uncertainty could imply a larger updating of growth expectations in response to the news. Thus it could lead to a larger cash flow channel, confounding my estimates. To verify this effect, I add macroeconomic uncertainty to regression (7). Table 10 contains results from regressions, where I add the interaction term between MNA surprise and macroeconomic uncertainty (MEU). MEU is a total macroeconomic uncertainty from Sydney Ludvigson's website. Controlling for MEU has three main effects. First, the coefficient on an interaction between MNA surprise and MU becomes slightly more negative. Seconds, the coefficient on Surprise*MEU is positive and highly significant. Third, the coefficient on MNA surprise loses its significance.
\paragraph{}
Overall, the results are consistent with the intuition that higher macroeconomic uncertainty implies a larger update of the expected cash flows after MNA. That's why the coefficient on Surprise*MEU is significantly positive. Moreover, a zero coefficient on the surprise term means that now MEU can fully capture a positive cash flow channel. Lastly, controlling for macroeconomic uncertainty slightly increases the estimated strength of the risk-free rate channel. This is consistent with a positive correlation between MU and MEU. Since the two channels go in the opposite direction, controlling for MEU allows measuring the risk-free rate channel more precisely, leading to its larger magnitude. 
\paragraph{}
My results and their interpretation rely on my ability to accurately proxy for MU using the implied volatility of 2-year Treasuries. As mentioned previously, I follow the literature in arguing that variation in 2-year T-notes reflects changes in monetary policy by capturing both short-term rate and forward guidance. It may be possible to validate my framework and its interpretation by looking at longer-maturity interest rates. For example, a 10-year rate is likely to have a large component related to economic growth, making it unsuitable for capturing time-series variation in a risk-free rate channel.
\paragraph{}
Table 11 describes stock response to the MNA surprise using different rate maturities to proxy for monetary uncertainty. Column 1 repeats the results from Table 7. The coefficient on the interaction term is negative and highly significant. Column 2 uses implied volatility of a 5-year rate. The coefficient on the interaction term loses its significance so the interaction term becomes redundant. Column 3 uses the implied volatility of the 10-year rate. Surprisingly, now the coefficient on the interaction term is significantly positive, while the coefficient on surprise loses its significance. This implies that now instead of proxying for a negative risk-free rate channel, an interaction term captures a positive cash flow channel. This is consistent with the idea that at longer maturities interest rates reflect economic growth rather than monetary policy. Overall, this result reinforces the interpretation of cash flow and risk-free rate channels of MNA response as well as validates the interpretation of the 2-year rate as a proxy for monetary policy.

\section{MNA surprise in the cross-section} \label{sec:Model}

This section uses surprises for 4 types of MNA to test whether these surprises are priced in the cross-section of expected returns. For each type of MNA, I use the same methodology. After estimating the covariance of stock returns with MNA surprise, I divide stocks into quintile portfolios sorted on this covariance and test whether there is a significant return spread.
\paragraph{}
As mentioned in Section 3, accurate identification of stock response to MNA requires using intraday data. Since individual stocks are less liquid than S\&P 500 and require more time to respond to the news, I use the window from 10 minutes before the news release to 2 hours after the announcement. Since small stocks are likely to respond to news slower due to low liquidity and resulting lags, I focus on constituents of the S\&P 500. To avoid effects due to inclusion/exclusion to S\&P 500 in the year t, I consider all stocks, which were constituents on S\&P500 at the end of year t-1.
\paragraph{}
I estimate covariances of individual stock returns and MNA surprises using a rolling window over 4 years. For each MNA and each stock I run regression (8):
\begin{equation}
   R_t = \alpha + \beta^{MNA} * Surprise_t + \beta*R^{SP500}_t + e_t.
\end{equation}
\paragraph{}
Each month I sort stocks into quintile portfolios, based on their $\beta^{MNA}$. I compute the returns of these decile portfolios and report them in Table 13. Tables 13 and 14 suggest that none of the 3 types of MNA are priced in the cross-section of expected stock returns. Long-short portfolios have returns between -10 and +21 bps with all t-statistics below 1. Abnormal return spreads of long-short portfolios are statistically and economically insignificant (-16 to + 3 bps).
\paragraph{}
One possible concern with this methodology is that I may be unable to get precise estimates of the sensitivities of individual stocks to MNA surprises. Figure 5 explores whether investors could implement these trading strategies in real time. To implement them, investors must be able to use their ex-ante estimates of $\beta^{MNA}$ to predict its realized values. In other words, pre-ranking $\beta^{MNA}$ must predict post-ranking $\beta^{MNA}$. I calculate post-ranking $\beta^{MNA}$ as a coefficient from regression (8) over a full sample for each quintile portfolio. Figure 5 shows that pre-ranking $\beta^{NFP}$ is a strong predictor of post-ranking $\beta^{NFP}$ with an almost perfectly linear relation between these variables. Results are similar for predictability of post-ranking $\beta^{PMI}$.
\paragraph{}
These results imply that risk premium is not affected by MNA and are consistent with time-series evidence, showing that response to MNA decreases at longer maturities. The concentration of the response to MNA in short-term and medium-term yields can mean that this news contains little information about a risk premium. Similarly, the lack of spread in the cross-section seems to imply that changes in key macroeconomic variables do not have a significant effect on risk premium.
For example, if a good MNA surprise implied a lower risk premium due to a lower probability of a rare disaster, then stocks with higher exposure to MNA surprise would be riskier and would have higher expected returns. While theoretically puzzling, my results are consistent with empirical research, which documents similar findings using a lower-frequency methodology (e.g., Herskovic, Moreira, and Muir, 2019). These results are consistent with asset pricing models, in which risk premium is not strongly correlated with the business cycle $^2$. 
\footnotetext[2]{One possible concern with this interpretation is that while my results suggest that the price of risk of the MNA factor is zero on average, it could vary in my sample, changing its sign. While this is hypothetically possible, I view this possibility as unlikely on theoretical grounds. I can not use empirical results to test this story, since the short sample period makes it impossible to separately estimate the price of MNA factor risk in separate subsamples with high precision.}

\section{Conclusion} \label{sec:Model}

This paper explores a large set of macroeconomic news announcements (MNA). The availability of new data enables improvements in identification, which allows me to obtain three important results. First, stocks exhibit a strong positive response to major MNA: 1 standard deviation of good surprise causes 11-25 bps higher returns. This response is the strongest for 4 commonly reported MNA: Non-farm payroll, PMI, Retail Sales, and Consumer Confidence. 
\paragraph{}
Second, I use monetary uncertainty to proxy for the strength of the monetary channel and decompose this response into cash flow and monetary components. 1 standard deviation of good MNA surprise leads to 30 bps returns from the cash flow channel and minus 23 bps per 1\% of monetary uncertainty from the monetary channel. The positive response of stocks to MNA via the cash flow channel is very large and amounts to 60\%-180\% of return variation during comparable no-news periods. Large monetary channel during periods of high monetary uncertainty masks a strong positive response of stocks to MNA, which explains why past research failed to detect this relation.
\paragraph{}
Third, I explore the response to major MNA in the cross-section of stock returns. Differential exposure to MNA shocks does not produce a spread in the cross-section of expected returns. This result is consistent with previous findings and suggests that the cash-flow channel and risk-free rate channel are the main channels through which MNAs affect stocks.

\pagebreak

\section{References:}
\begin{enumerate}
    \item{Andersen, Torben G., Tim Bollerslev, Francis X. Diebold, and Clara Vega. "Micro effects of macro announcements: Real-time price discovery in foreign exchange." American Economic Review 93, no. 1 (2003): 38-62.}
    \item{Ang, Andrew, Geert Bekaert, and Min Wei. "Do macro variables, asset markets, or surveys forecast inflation better?." Journal of Monetary Economics 54, no. 4 (2007): 1163-1212.}
    \item{Ball, Ray, and Philip Brown. "An empirical evaluation of accounting income numbers." Journal of accounting research (1968): 159-178.}
    \item{Bernanke, Ben and Kuttner, Kenneth (2005), What Explains the Stock Market's Reaction to Federal Reserve Policy?. The Journal of Finance, 60: 1221-1257.}
    \item{Boyd, John H., Jian Hu, and Ravi Jagannathan. "The stock market's reaction to unemployment news: Why bad news is usually good for stocks." The Journal of Finance 60, no. 2 (2005): 649-672.}
    \item{Cutler, David M., James M. Poterba, and Lawrence H. Summers. What moves stock prices?. No. w2538. National Bureau of Economic Research, 1988.}
    \item{Cieslak, Anna, Adair Morse, and Annette Vissing‐Jorgensen. "Stock returns over the FOMC cycle." The Journal of Finance 74, no. 5 (2019): 2201-2248.}
    \item{Ernst, Rory, Thomas Gilbert, and Christopher M. Hrdlicka. "More than 100\% of the equity premium: How much is really earned on macroeconomic announcement days?." Available at SSRN 3469703 (2019).}
    \item{Fama, Eugene F., Lawrence Fisher, Michael C. Jensen, and Richard Roll. "The adjustment of stock prices to new information." International economic review 10, no. 1 (1969): 1-21.}
    \item{Flannery, Mark J., and Aris A. Protopapadakis. "Macroeconomic factors do influence aggregate stock returns." The review of financial studies 15, no. 3 (2002): 751-782.}
    \item{Gertler, Mark, and Peter Karadi. "Monetary policy surprises, credit costs, and economic activity." American Economic Journal: Macroeconomics 7, no. 1 (2015): 44-76.}
    \item{Gürkaynak, Refet S., Brian P. Sack, and Eric T. Swanson. "Do actions speak louder than words? The response of asset prices to monetary policy actions and statements." The Response of Asset Prices to Monetary Policy Actions and Statements (November 2004) (2004).}
    \item{Hanson, Samuel G., and Jeremy C. Stein. "Monetary policy and long-term real rates." Journal of Financial Economics 115, no. 3 (2015): 429-448.}
    \item{Herskovic, Bernard, Alan Moreira, and Tyler Muir. "Hedging risk factors." (2019).}
    \item{Hu, Xing, Jun Pan, Jiang Wang, and Haoxiang Zhu. Premium for heightened uncertainty: Solving the fomc puzzle. National Bureau of Economic Research, 2019.}
    \item{Laarits, Toomas. "Pre-announcement risk." NYU Stern School of Business (2019).}
    \item{Law, Tzuo Hann, Dongho Song, and Amir Yaron. "Fearing the fed: How wall street reads main street." Available at SSRN 3092629 (2021).}
    \item{Lucca, David O., and Emanuel Moench. "The pre‐FOMC announcement drift." The Journal of finance 70, no. 1 (2015): 329-371.}
    \item{McQueen, Grant, and V. Vance Roley. "Stock prices, news, and business conditions." The review of financial studies 6, no. 3 (1993): 683-707.}
    \item{Nakamura, Emi, and Jón Steinsson. "High-frequency identification of monetary non-neutrality: the information effect." The Quarterly Journal of Economics 133, no. 3 (2018): 1283-1330.}
    \item{Pearce, Douglas K., and V. Vance Roley. Stock prices and economic news. No. w1296. National bureau of economic research, 1984.}
    \item{Rigobon, Roberto, and Brian Sack. "The impact of monetary policy on asset prices." Journal of monetary economics 51, no. 8 (2004): 1553-1575.}
    \item{Rossi, Barbara, and Tatevik Sekhposyan. "Macroeconomic uncertainty indices based on nowcast and forecast error distributions." American Economic Review 105, no. 5 (2015): 650-55.}
    \item{Savor, Pavel, and Mungo Wilson. "How much do investors care about macroeconomic risk? Evidence from scheduled economic announcements." Journal of Financial and Quantitative Analysis (2013): 343-375.}
    \item{Savor, Pavel, and Mungo Wilson. "Asset pricing: A tale of two days." Journal of Financial Economics 113, no. 2 (2014): 171-201.}
    \item{Schwert, G. William. "The adjustment of stock prices to information about inflation." The Journal of Finance 36, no. 1 (1981): 15-29.}
    \item{Swanson, Eric T., and John C. Williams. "Measuring the effect of the zero lower bound on yields and exchange rates in the U.K. and Germany." Journal of International Economics 92 (2013):  S2-S21.}
    \item{Swanson, Eric T., and John C. Williams. "Measuring the effect of the zero lower bound on medium-and longer-term interest rates." American economic review 104, no. 10 (2014): 3154-85.}
    \item{Vissing-Jorgensen, Annette. "Central banking with many voices: The communications arms race." (2019).}
    \item{Zhou, John Cong. Asset Price Reactions to News at Zero Lower Bound. working paper, 2014}
\end{enumerate}
    
\pagebreak

\section{Appendix A: Empirical Results} \label{sec:Model}

\begin{figure}[h!]
\textbf{Figure 1: Implied volatility of 2-year Treasuries}
\vskip 6 pt
\begin{flushleft}
{The figure describes the implied volatility of the yield on 2-year Treasury notes in percentage points. To proxy for this implied volatility, I use Black implied volatility of swaptions with a 1-year maturity and 2-year tenor. In other words, I use options on interest rate swaps with a length of two years and an expiration of 1 year. Since Black volatility measures movement per 1\% of interest rate, I rescale this implied volatility by the level of the 2-year Treasury rate. For example, the implied volatility of 100 bps in July 2002 is a product of the swaption implied volatility of 42\% and 2.4\% level of the 2-year Treasury rate. The red line shows the evolution of the Fed Funds Rate over the same sample period.}
\end{flushleft}
\centering
\vspace{0.64cm}
\includegraphics[width=1\textwidth]{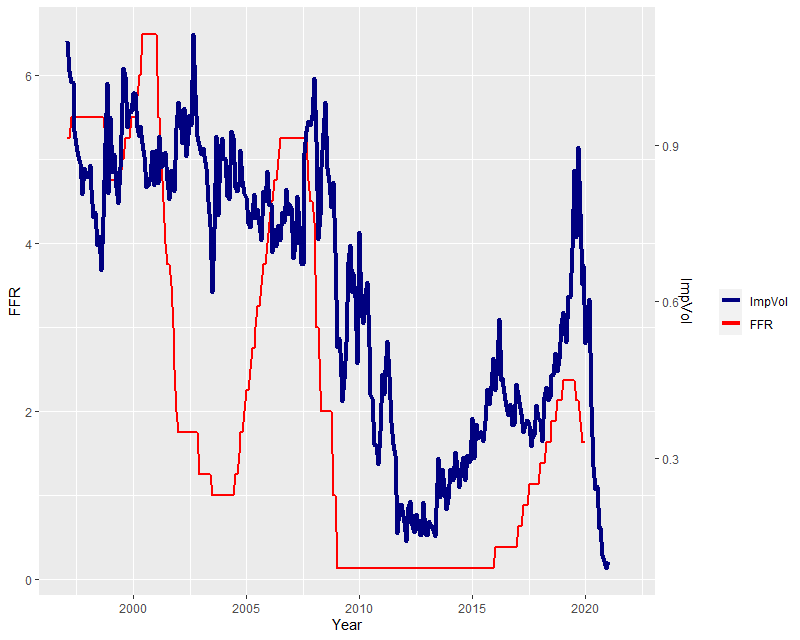}

\end{figure}

\begin{figure}[h!]
\textbf{Figure 2: Aggregate price dynamics around MNA}
\vskip 6 pt
\begin{flushleft}
{The figure describes the evolution of S\&P500 futures prices around 5 main MNA. I divide all MNA surprises into quartiles based on the size of the surprise. Quartile 1 corresponds to the most negative surprises, while Quartile 4 corresponds to the most positive surprises. The plot shows the evolution of S\&P500 futures at a 1-minute frequency. Y-axis describes the return relative to the price level 1 minute before the announcement. }
\end{flushleft}
\centering
\vspace{0.64cm}
\includegraphics[width=1\textwidth]{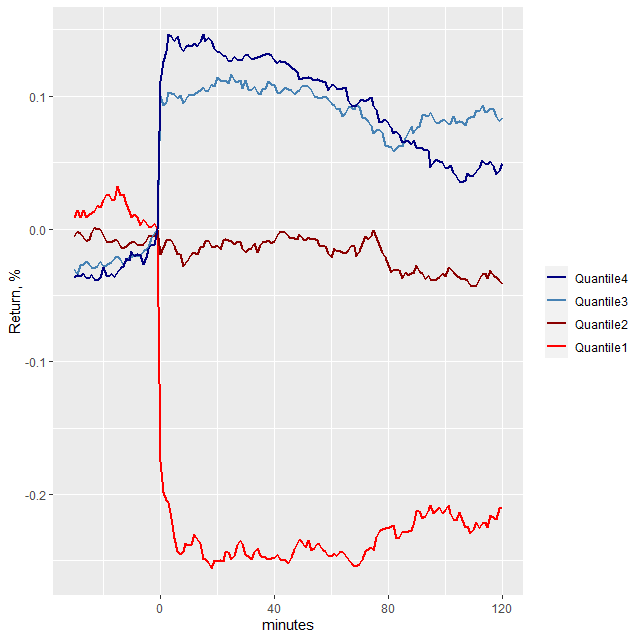}
\end{figure}

\begin{figure}[h!]
\textbf{Figure 3: Term structure response to MNA surprise}
\vskip 6 pt
\begin{flushleft}
{The figure describes the response of interest rates to MNA surprise across maturities. Y-axis depicts coefficient estimate b from daily-frequency regression $\Delta i_t = a + b Surprise_t + \epsilon_t.$ The sample includes the most important MNA: Non-farm payrolls, PMI, Retail sales, and Consumer sentiment.}
\end{flushleft}
\centering
\vspace{0.64cm}
\includegraphics[width=0.9\textwidth]{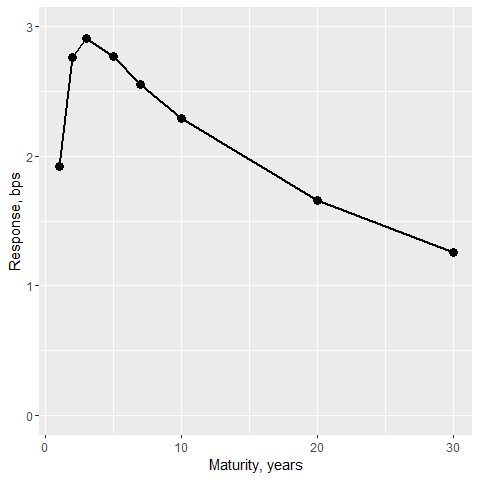}
\end{figure}

\begin{figure}[h!]
\textbf{Figure 4: Extrapolation of the stock response to MNA to the earlier sample}
\vskip 6 pt
\begin{flushleft}
{The figure describes the projected response of stocks to MNA surprise over 1978-2019.
I evaluate this response by estimating it in the 1997-2019 sample and extrapolating these results to the earlier sample. I estimate the coefficients $b, d$ from the regression $R_t = a + b Surprise_t + c MU^{rv}_{t-1} + d Surprise_t MU^{rv}_{t-1} + \epsilon_t$ to be 20 and -16 bps respectively. I extrapolate the results to the earlier sample, assuming a constant cash flow channel of the size $b$, and a time-varying risk-free rate channel equal to $d MU^{rv}_{t-1}$. The green line shows a positive cash flow channel. The red line shows a negative risk-free rate channel. The total response is a shaded area, which is green when the response is positive and red when it is negative. I use realized volatility of 2-year Treasury notes to calculate a time-varying risk-free channel. The sample includes the most important MNA: Non-farm payrolls, PMI, Retail sales, and Consumer sentiment.}
\end{flushleft}
\centering
\vspace{0.64cm}
\includegraphics[width=0.9\textwidth]{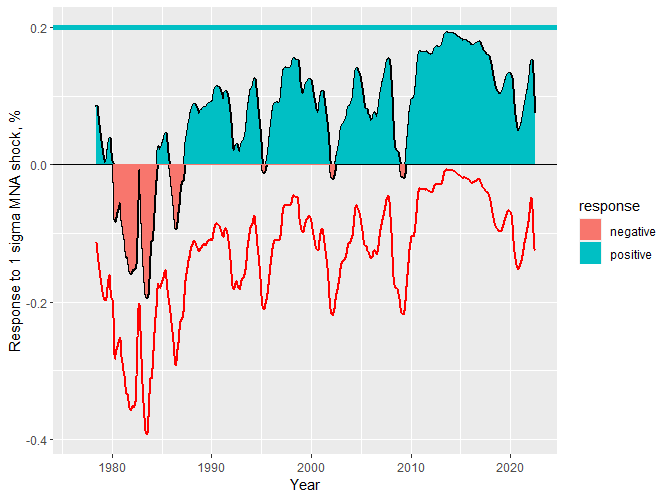}
\end{figure}

\begin{figure}[h!]
\textbf{Figure 5: Pre-ranking and post-ranking betas}
\vskip 6 pt
\begin{flushleft}
{The figure describes pre-ranking and post-ranking betas of quintile portfolios, sorted on pre-ranking beta. For each panel, pre-ranking beta $\beta_{MNA}$ is a coefficient from the regression of stock returns on MNA surprise for a certain type of MNA. I estimate post-ranking $\beta_{MNA}$ using a full sample for each quintile value-weighted portfolio, formed on pre-ranking $\beta_{MNA}$. }
\end{flushleft}
\centering
\vspace{0.64cm}
\includegraphics[width=0.9\textwidth]{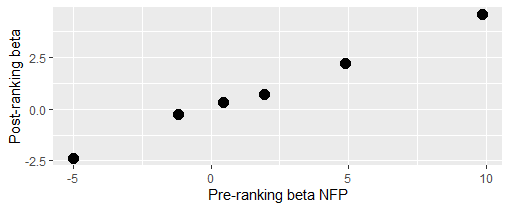}
\vspace{0.64cm}
\includegraphics[width=0.9\textwidth]{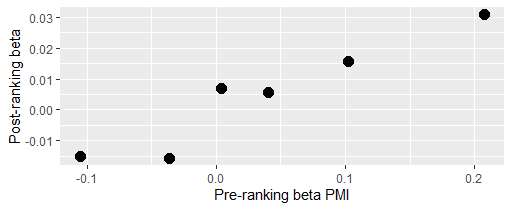}
\end{figure}

\pagebreak

\newpage

\clearpage

% keep old table as of 01/08/22
% old versions of tables are possibly produced on 3/22 or 3/23 of 2021.
\begin{table}[!htbp] \centering 
  \caption{\textbf{Macroeconomic News Announcements}} 
  \label{} 
  \begin{flushleft}
    {\medskip\small
 The table reports details about all MNA at monthly frequency. N stands for the number of observations in the 1997-2019 sample, while Day means a business day of the month when the news is released.}
    \medskip
    \end{flushleft}
\begin{tabular}{@{\extracolsep{5pt}} llll} 
\\[-1.8ex]\hline 
\hline \\[-1.8ex] 
News & N & Time & Day \\ 
\hline \\[-1.8ex] 
Non-Farm Payroll & $276$ & 8:30 & 1-5 \\ 
ISM Manufacturing & $278$ & 10:00 & 1-2 \\ 
Retail Sales Advance MoM & $277$ & 08:30 & 9-12 \\ 
Construction Spending MoM & $272$ & 10:00 & 1-2 \\ 
Conf. Board Consumer Confidence & $275$ & 10:00 & 18-21 \\ 
Capacity Utilization & $276$ & 09:15 & 10-13 \\ 
U. of Mich. Sentiment F & $248$ & 10:00 & 1-2, 16-21 \\ 
Trade Balance & $277$ & 08:30 & 3-15 \\ 
Business Inventories & $270$ & 10:00 & 9-12 \\ 
Housing Starts & $263$ & 08:30 & 10-14 \\ 
Factory Orders & $279$ & 10:00 & 2-4 \\ 
Leading Index & $274$ & 10:00 & 1-3, 10-16 \\ 
U. of Mich. Sentiment P & $247$ & 10:00 & 8-13 \\ 
Monthly Budget Statement & $278$ & 14:00 & 8-17 \\ 
New Home Sales & $268$ & 10:00 & 16-21 \\ 
Durable Goods Orders & $258$ & 08:30 & 16-20 \\ 
Consumer Credit & $278$ & 15:00 & 5 \\ 
\hline \\[-1.8ex] 
\end{tabular} 
\end{table}

% keep old table as of 01/08/22
\begin{table}[!htbp] \centering 
  \caption{\textbf{Absolute returns during MNA}} 
  \label{} 
    \begin{flushleft}
    {\medskip\small
 The table reports an absolute value of returns of S\&P 500 futures over a 30-minutes window around MNA. I test the hypothesis that the absolute value of returns is higher during the MNA window than during a similar window on days without MNA. Mean absolute returns on non-MNA days range between 14 and 24 bps depending on the time. The last two columns report p-values from the two tests: the T-test and the Mann-Whitney-Wilcoxon test. The null hypothesis is that mean absolute returns during MNA periods are equal to mean absolute returns over comparable periods without MNA.}
    \medskip
    \end{flushleft}
\begin{tabular}{@{\extracolsep{5pt}} lllll} 
\\[-1.8ex]\hline 
\hline \\[-1.8ex] 
News & Mean & Median & p-value (t) & p-value (MWW) \\ 
\hline \\[-1.8ex] 
Non-Farm Payroll & $0.41$ & $0.29$ & $0$ & $0$ \\ 
ISM Manufacturing & $0.36$ & $0.27$ & $0.0000$ & $0$ \\ 
Retail Sales Advance MoM & $0.21$ & $0.13$ & $0.0000$ & $0$ \\ 
Construction Spending MoM & $0.34$ & $0.23$ & $0.0000$ & $0$ \\ 
Conf. Board Consumer Confidence & $0.29$ & $0.18$ & $0.03$ & $0.12$ \\ 
Capacity Utilization & $0.17$ & $0.10$ & $0.14$ & $0.37$ \\ 
U. of Mich. Sentiment F & $0.27$ & $0.18$ & $0.28$ & $0.17$ \\ 
Trade Balance & $0.16$ & $0.10$ & $0.31$ & $0.03$ \\ 
Business Inventories & $0.23$ & $0.16$ & $0.40$ & $0.53$ \\ 
Housing Starts & $0.15$ & $0.09$ & $0.40$ & $0.07$ \\ 
Factory Orders & $0.24$ & $0.17$ & $0.46$ & $0.97$ \\ 
Leading Index & $0.26$ & $0.19$ & $0.47$ & $0.29$ \\ 
U. of Mich. Sentiment P & $0.26$ & $0.16$ & $0.76$ & $0.66$ \\ 
Monthly Budget Statement & $0.17$ & $0.11$ & $0.78$ & $0.33$ \\ 
New Home Sales & $0.25$ & $0.16$ & $0.79$ & $0.86$ \\ 
Durable Goods Orders & $0.14$ & $0.09$ & $0.84$ & $0.17$ \\ 
Consumer Credit & $0.20$ & $0.12$ & $0.91$ & $0.66$ \\ 
\hline \\[-1.8ex] 
\end{tabular} 
\end{table}

% keep old table as of 01/08/22
\begin{table}[!htbp] \centering 
  \caption{\textbf{Summary statistics of returns and announcement surprises}} 
  \label{}
  \begin{flushleft}
    {\medskip\small
 The table reports summary statistics of macroeconomic variables, announcement surprises and announcement returns (SP500) for each major MNA. px stands for $x^{th}$ percentile. Returns are reported in percentage points, surprises are normalized to have a standard deviation of 1.}
    \medskip
    \end{flushleft}
\begin{tabular}{@{\extracolsep{-3pt}} lllllllllllll} 
\\[-1.8ex]\hline 
\hline \\[-1.8ex] 
News & Variable & Min & p1 & p10 & p25 & Median & p75 & p90 & p99 & Max & Mean & SD \\ 
\hline \\[-1.8ex] 
 & CFNAI & 3.35 & -2.68 & -0.61 & -0.31 & 0.02 & 0.28 & 0.50 & 0.94 & 1.21 & -0.07 & 0.60 \\ 
 & USREC & 0 & 0 & 0 & 0 & 0 & 0 & 0 & 1 & 1 & 0.09 & 0.29 \\ 
 & FFR\_{t-1} & 0.12 & 0.12 & 0.12 & 0.12 & 1.62 & 4.75 & 5.50 & 6.50 & 6.50 & 2.25 & 2.14 \\ 
 & $\Delta$ rate & -0.09 & -0.05 & -0.01 & 0 & 0 & 0 & 0.01 & 0.04 & 0.09 & 0 & 0.01 \\ 
NFP & SP500 & -2.03 & -1.46 & -0.66 & -0.20 & 0.10 & 0.31 & 0.72 & 1.41 & 1.97 & 0.04 & 0.56 \\ 
NFP & surprise & -3.93 & -2.90 & -1.31 & -0.68 & -0.06 & 0.44 & 1.02 & 2.19 & 3.22 & -0.13 & 1 \\ 
PMI & SP500 & -3.26 & -1.34 & -0.57 & -0.29 & 0.01 & 0.23 & 0.50 & 1.47 & 2.17 & 0 & 0.52 \\ 
PMI & surprise & -3.23 & -2.45 & -1.13 & -0.59 & 0 & 0.66 & 1.32 & 2.37 & 3.98 & 0.06 & 1 \\ 
Retail & SP500 & -1.49 & -0.93 & -0.30 & -0.08 & 0.02 & 0.16 & 0.33 & 0.93 & 1.04 & 0.02 & 0.32 \\ 
Retail & surprise & -3 & -2.62 & -0.94 & -0.56 & 0 & 0.37 & 0.94 & 2.68 & 8.62 & -0.02 & 1 \\ 
Constr & SP500 & -3.26 & -1.23 & -0.47 & -0.24 & 0.03 & 0.22 & 0.50 & 1.45 & 2.17 & 0.02 & 0.50 \\ 
Constr & surprise & -7.71 & -2.18 & -1.09 & -0.51 & 0 & 0.45 & 0.86 & 2.17 & 4.20 & -0.06 & 1 \\ 
Unempl & SP500 & -2.03 & -1.46 & -0.66 & -0.20 & 0.10 & 0.31 & 0.72 & 1.41 & 1.97 & 0.04 & 0.56 \\ 
Unempl & surprise & -3.52 & -2.34 & -1.41 & -0.70 & 0 & 0.70 & 0.70 & 2.11 & 2.81 & -0.18 & 1 \\ 
\hline \\[-1.8ex] 
\end{tabular}
\end{table}

\begin{table}[!htbp] \centering 
  \caption{\textbf{Response of S\&P500 to MNA, $R_t = a + b Surprise_t + \epsilon_t.$}} 
  \label{}
  \begin{flushleft}
    {\medskip\small
 The table reports the response of S\&P 500 futures to MNA surprises over the 30-minutes announcement window. Each row reports results for different MNA. $R^{2}$ is adjusted $R^{2}$ from this regression.}
    \medskip
    \end{flushleft}
\begin{tabular}{@{\extracolsep{5pt}} llll} 
\\[-1.8ex]\hline 
\hline \\[-1.8ex] 
Event & b & T-statistic(b) & $R^2$ \\ 
\hline \\[-1.8ex] 
Change in Nonfarm Payrolls & $0.25$ & $7.82$ & $0.20$ \\ 
ISM Manufacturing & $0.16$ & $5.20$ & $0.09$ \\ 
Conf. Board Consumer Confidence & $0.11$ & $4.87$ & $0.09$ \\ 
Retail Sales Advance MoM & $0.11$ & $5.87$ & $0.12$ \\ 
Construction Spending MoM & $0.07$ & $2.44$ & $0.02$ \\ 
Unemployment Rate & $0.06$ & $1.60$ & $0.01$ \\ 
\hline \\[-1.8ex] 
\end{tabular} 
\end{table}

% keep old table as of 01/08/22
\begin{table}[!htbp] \centering 
  \caption{\textbf{Response of SP500 to PMI news announcement}} 
  \label{}
  \begin{flushleft}
    {\medskip\small
 The table documents the response of S\&P 500 futures to PMI news surprises. Each column reports estimates from the regression $R_t = a + b Surprise_t + \epsilon_t.$ for different window widths over which returns are measured. The first column reports regression, where the dependent variable is a daily return of the CRSP value-weighted index. The dependent variable in the second column is S\&P 500 futures at a daily frequency. The third column uses S\&P 500 futures returns between the previous day's close (4 pm) and 20 minutes after the news release. The last column reports estimates from the main specification with a 30-minutes window and S\&P 500 futures.}
    \medskip
    \end{flushleft}
\begin{tabular}{@{\extracolsep{5pt}}lcccc} 
\\[-1.8ex]\hline 
\hline \\[-1.8ex] 
 & \multicolumn{4}{c}{\textit{Dependent variable:}} \\ 
\cline{2-5} 
\\[-1.8ex] & vwretd & SP500 & SP500 & SP500 \\ 
\\[-1.8ex] & (1) & (2) & (3) & (4)\\ 
\hline \\[-1.8ex] 
 PMI surprise & 0.160$^{**}$ & 0.169$^{**}$ & 0.193$^{***}$ & 0.156$^{***}$ \\ 
  & [1.994] & [2.027] & [3.142] & [5.135] \\ 
  & & & & \\ 
 Constant & 0.163$^{**}$ & 0.165$^{*}$ & 0.105$^{*}$ & $-$0.008 \\ 
  & [2.018] & [1.961] & [1.696] & [$-$0.260] \\ 
  & & & & \\ 
\hline \\[-1.8ex] 
Announcement window & Day & Day & Half-day & 30 minutes \\ 
Observations & 276 & 266 & 266 & 266 \\ 
Adjusted R$^{2}$ & 0.011 & 0.012 & 0.032 & 0.087 \\ 
\hline 
\hline \\[-1.8ex] 
\textit{Note:}  & \multicolumn{4}{r}{$^{*}$p$<$0.1; $^{**}$p$<$0.05; $^{***}$p$<$0.01} \\ 
\end{tabular} 
\end{table}

% Table created by stargazer v.5.2.2 by Marek Hlavac, Harvard University. E-mail: hlavac at fas.harvard.edu
% Date and time: Mon, Mar 22, 2021 - 8:35:10 PM
\begin{table}[!htbp] \centering 
  \caption{\textbf{Stock response to major MNA}} 
  \label{} 
  \begin{flushleft}
    {\medskip\small
 The table documents the response of S\&P 500 futures to MNA surprises among the 4 most important MNA: NFP, PMI, Consumer Confidence, and Retail. Each column reports estimates from the regression $R_t = a + b Surprise_t + \epsilon_t.$ for different subsamples: periods of recession or expansion, half-samples with low/high levels of monetary uncertainty.}
    \medskip
    \end{flushleft}
\begin{tabular}{@{\extracolsep{5pt}}lcccc} 
\\[-1.8ex]\hline 
\hline \\[-1.8ex] 
 & \multicolumn{4}{c}{\textit{Dependent variable:}} \\ 
\cline{2-5} 
\\[-1.8ex] & \multicolumn{4}{c}{SP500} \\ 
\\[-1.8ex] & (1) & (2) & (3) & (4)\\ 
\hline \\[-1.8ex] 
 Surprise & 0.203$^{***}$ & 0.141$^{***}$ & 0.244$^{***}$ & 0.094$^{***}$ \\ 
  & [4.552] & [9.998] & [12.869] & [4.987] \\ 
  & & & & \\ 
 Constant & $-$0.030 & 0.013 & 0.006 & 0.004 \\ 
  & [$-$0.444] & [1.014] & [0.384] & [0.206] \\ 
  & & & & \\ 
\hline \\[-1.8ex] 
Subsample & Recession & Expansion & Low mon. uncert. & High mon. uncert. \\ 
Observations & 96 & 891 & 514 & 473 \\ 
Adjusted R$^{2}$ & 0.172 & 0.100 & 0.243 & 0.048 \\ 
\hline 
\hline \\[-1.8ex] 
\textit{Note:}  & \multicolumn{4}{r}{$^{*}$p$<$0.1; $^{**}$p$<$0.05; $^{***}$p$<$0.01} \\ 
\end{tabular} 
\end{table}

% being updated
\begin{table}[!htbp] \centering 
  \caption{\textbf{Response to major MNA with interaction terms}} 
  \label{}
    \begin{flushleft}
    {\medskip\small
 The table documents the response of stocks and 2-year Treasury rates to 4 major MNA surprises: $R_t = a + b Surprise_t + c MU_{t-1} + d Surprise_t MU_{t-1} + \epsilon_t.$ The first three columns document response of the S\&P 500, and the last two report the response of the 2-year Treasury rate. All responses are measured over a 30-minute window around an announcement. MU stands for monetary uncertainty, i.e., implied volatility of the 2-year Treasury rate. CFNAI is Chicago Fed National Activity Index.}
    \medskip
    \end{flushleft}
\begin{tabular}{@{\extracolsep{5pt}}lccccc} 
\\[-1.8ex]\hline 
\hline \\[-1.8ex] 
 & \multicolumn{5}{c}{\textit{Dependent variable:}} \\ 
\cline{2-6} 
\\[-1.8ex] & \multicolumn{3}{c}{SP500} & \multicolumn{2}{c}{$\Delta$ rate} \\ 
\\[-1.8ex] & (1) & (2) & (3) & (4) & (5)\\ 
\hline \\[-1.8ex] 
 Surprise & 0.15$^{***}$ & 0.30$^{***}$ & 0.29$^{***}$ & 2.16$^{***}$ & 0.17 \\ 
  & [11.29] & [7.82] & [7.35] & [17.35] & [0.47] \\ 
  & & & & & \\ 
 MU &  & $-$0.01 & $-$0.02 &  & 0.46 \\ 
  &  & [$-$0.29] & [$-$0.29] &  & [0.98] \\ 
  & & & & & \\ 
 CFNAI &  &  & $-$0.02 &  &  \\ 
  &  &  & [$-$0.66] &  &  \\ 
  & & & & & \\ 
 Surprise*MU &  & $-$0.23$^{***}$ & $-$0.22$^{***}$ &  & 2.93$^{***}$ \\ 
  &  & [$-$4.18] & [$-$4.08] &  & [5.86] \\ 
  & & & & & \\ 
 Surprise*CFNAI &  &  & $-$0.05$^{***}$ &  &  \\ 
  &  &  & [$-$2.96] &  &  \\ 
  & & & & & \\ 
 Constant & 0.01 & 0.02 & 0.02 & $-$0.01 & $-$0.28 \\ 
  & [0.96] & [0.59] & [0.63] & [$-$0.08] & [$-$0.90] \\ 
  & & & & & \\ 
\hline \\[-1.8ex] 
Observations & 978 & 978 & 978 & 1,008 & 1,005 \\ 
Adjusted R$^{2}$ & 0.11 & 0.13 & 0.13 & 0.23 & 0.25 \\ 
\hline 
\hline \\[-1.8ex] 
\textit{Note:}  & \multicolumn{5}{r}{$^{*}$p$<$0.1; $^{**}$p$<$0.05; $^{***}$p$<$0.01} \\ 
\end{tabular} 
\end{table}

\begin{table}[!htbp] \centering 
  \caption{\textbf{Interest rate response to major MNA}} 
  \label{}
    \begin{flushleft}
    {\medskip\small
 The table documents the response of interest rates at various maturities to MNA's surprise. The table report results from the following regression at daily frequency: $\Delta i_t = a + b Surprise_t + \epsilon_t$. Dependent variables are changes in daily treasury yields at maturities from 1 year to 30 years.}
    \medskip
    \end{flushleft}
\begin{tabular}{@{\extracolsep{2pt}}lcccccccc} 
\\[-1.8ex]\hline 
\hline \\[-1.8ex] 
 & \multicolumn{8}{c}{\textit{Dependent variable:}} \\ 
\cline{2-9} 
\\[-1.8ex] & $\Delta$ i\_1 & $\Delta$ i\_2 & $\Delta$ i\_3 & $\Delta$ i\_5 & $\Delta$ i\_7 & $\Delta$ i\_{10} & $\Delta$ i\_{20} & $\Delta$ i\_{30} \\ 
\\[-1.8ex] & (1) & (2) & (3) & (4) & (5) & (6) & (7) & (8)\\ 
\hline \\[-1.8ex] 
 Surprise & 1.92$^{***}$ & 2.77$^{***}$ & 2.91$^{***}$ & 2.77$^{***}$ & 2.56$^{***}$ & 2.29$^{***}$ & 1.66$^{***}$ & 1.26$^{***}$ \\ 
  & [9.42] & [9.85] & [9.92] & [9.16] & [8.78] & [8.47] & [7.62] & [7.09] \\ 
  & & & & & & & & \\ 
 Constant & $-$0.07 & $-$0.07 & 0.10 & 0.31 & 0.39 & 0.37 & 0.25 & 0.19 \\ 
  & [$-$0.30] & [$-$0.24] & [0.31] & [0.91] & [1.20] & [1.24] & [1.05] & [0.94] \\ 
  & & & & & & & & \\ 
\hline \\[-1.8ex] 
Observations & 491 & 491 & 491 & 491 & 491 & 491 & 491 & 491 \\ 
Adjusted R$^{2}$ & 0.15 & 0.16 & 0.17 & 0.14 & 0.13 & 0.13 & 0.10 & 0.09 \\ 
\hline 
\hline \\[-1.8ex] 
\textit{Note:}  & \multicolumn{8}{r}{$^{*}$p$<$0.1; $^{**}$p$<$0.05; $^{***}$p$<$0.01} \\ 
\end{tabular} 
\end{table}

% Table created by stargazer v.5.2.3 by Marek Hlavac, Social Policy Institute. E-mail: marek.hlavac at gmail.com
% Date and time: Sat, Oct 08, 2022 - 10:24:34 AM
\begin{table}[!htbp] \centering 
  \caption{\textbf{Response to major MNA with interaction terms, realized volatility}} 
  \label{}
    \begin{flushleft}
    {\medskip\small
 The table documents response of S\&P 500 to 4 major MNA surprises: $R_t = a + b Surprise_t + c MU^{rv}_{t-1} + d Surprise_t MU^{rv}_{t-1} + \epsilon_t.$ All responses are measured over a 30-minutes window around an announcement. $MU^{rv}$ stands for monetary uncertainty. In this regression, I proxy for MU using realized volatility of 2-year Treasury yield, $MU^{rv}$. The sample includes the most important MNA: Non-farm payrolls, PMI, Retail sales, and Consumer sentiment.}
    \medskip
    \end{flushleft}
\begin{tabular}{@{\extracolsep{5pt}}lcc} 
\\[-1.8ex]\hline 
\hline \\[-1.8ex] 
 & \multicolumn{2}{c}{\textit{Dependent variable:}} \\ 
\cline{2-3} 
\\[-1.8ex] & \multicolumn{2}{c}{SP500} \\ 
\\[-1.8ex] & (1) & (2)\\ 
\hline \\[-1.8ex] 
 Surprise & 0.15$^{***}$ & 0.20$^{***}$ \\ 
  & [11.29] & [7.95] \\ 
  & & \\ 
 $MU^{rv}$ &  & $-$0.003 \\ 
  &  & [$-$0.04] \\ 
  & & \\ 
 CFNAI &  & $-$0.02 \\ 
  &  & [$-$0.80] \\ 
  & & \\ 
 $Surprise*MU^{rv}$ &  & $-$0.16$^{***}$ \\ 
  &  & [$-$2.94] \\ 
  & & \\ 
 Surprise*CFNAI &  & $-$0.09$^{***}$ \\ 
  &  & [$-$4.12] \\ 
  & & \\ 
 Constant & 0.01 & 0.01 \\ 
  & [0.96] & [0.65] \\ 
  & & \\ 
\hline \\[-1.8ex] 
Observations & 978 & 978 \\ 
Adjusted R$^{2}$ & 0.11 & 0.13 \\ 
\hline 
\hline \\[-1.8ex] 
\textit{Note:}  & \multicolumn{2}{r}{$^{*}$p$<$0.1; $^{**}$p$<$0.05; $^{***}$p$<$0.01} \\ 
\end{tabular} 
\end{table}

% Table created by stargazer v.5.2.3 by Marek Hlavac, Social Policy Institute. E-mail: marek.hlavac at gmail.com
% Date and time: Sat, Oct 08, 2022 - 3:08:35 PM
\begin{table}[!htbp] \centering 
  \caption{\textbf{Response to major MNA with interaction terms, macroeconomic uncertainty}} 
  \label{}
    \begin{flushleft}
    {\medskip\small
 The table documents response of S\&P500 to 4 major MNA surprises: $R_t = a + b Surprise_t + c MU_{t-1} + d Surprise_t MU_{t-1} + \epsilon_t.$ All responses are measured over a 30-minute window around the announcement. MU stands for monetary uncertainty, i.e., implied volatility of the 2-year Treasury rate. MEU stands for macroeconomic uncertainty and is taken from the website of Sydney Ludvigson.}
    \medskip
    \end{flushleft}
\begin{tabular}{@{\extracolsep{5pt}}lcccc} 
\\[-1.8ex]\hline 
\hline \\[-1.8ex] 
 & \multicolumn{4}{c}{\textit{Dependent variable:}} \\ 
\cline{2-5} 
\\[-1.8ex] & \multicolumn{4}{c}{SP500} \\ 
\\[-1.8ex] & (1) & (2) & (3) & (4)\\ 
\hline \\[-1.8ex] 
 Surprise & 0.15$^{***}$ & 0.31$^{***}$ & 0.29$^{***}$ & $-$0.03 \\ 
  & [11.29] & [8.11] & [7.74] & [$-$0.31] \\ 
  & & & & \\ 
 MU &  & 0.02 & 0.02 & 0.04 \\ 
  &  & [0.31] & [0.33] & [0.83] \\ 
  & & & & \\ 
 CFNAI &  &  & $-$0.02 & $-$0.05$^{*}$ \\ 
  &  &  & [$-$0.70] & [$-$1.71] \\ 
  & & & & \\ 
 MEU &  &  &  & $-$0.33 \\ 
  &  &  &  & [$-$1.64] \\ 
  & & & & \\ 
 Surprise*MU &  & $-$0.23$^{***}$ & $-$0.23$^{***}$ & $-$0.27$^{***}$ \\ 
  &  & [$-$4.41] & [$-$4.40] & [$-$5.06] \\ 
  & & & & \\ 
 Surprise*CFNAI &  &  & $-$0.06$^{***}$ & 0.02 \\ 
  &  &  & [$-$3.09] & [0.67] \\ 
  & & & & \\ 
 Surprise*MEU &  &  &  & 0.55$^{***}$ \\ 
  &  &  &  & [3.22] \\ 
  & & & & \\ 
 Constant & 0.01 & 0.003 & 0.004 & 0.19 \\ 
  & [0.96] & [0.09] & [0.11] & [1.61] \\ 
  & & & & \\ 
\hline \\[-1.8ex] 
Observations & 978 & 978 & 978 & 978 \\ 
Adjusted R$^{2}$ & 0.11 & 0.13 & 0.14 & 0.15 \\ 
\hline 
\hline \\[-1.8ex] 
\textit{Note:}  & \multicolumn{4}{r}{$^{*}$p$<$0.1; $^{**}$p$<$0.05; $^{***}$p$<$0.01} \\ 
\end{tabular} 
\end{table}

% Table created by stargazer v.5.2.3 by Marek Hlavac, Social Policy Institute. E-mail: marek.hlavac at gmail.com
% Date and time: Sat, Oct 08, 2022 - 6:28:36 PM
\begin{table}[!htbp] \centering 
  \caption{\textbf{Stock response to major MNA with interaction terms using MU constructed from different T-note maturities}} 
  \label{}
    \begin{flushleft}
    {\medskip\small
 The table documents response of S\&P500 to 4 major MNA surprises: $R_t = a + b Surprise_t + c MU_{t-1} + d Surprise_t MU_{t-1} + \epsilon_t.$ All responses are measured over a 30-minutes window around the announcement. I use 2, 5, and 10-year Treasuries to try to proxy for monetary uncertainty. Imv2, Imv5, and Imv10 are such implied volatilities.}
    \medskip
    \end{flushleft}
\begin{tabular}{@{\extracolsep{5pt}}lccc} 
\\[-1.8ex]\hline 
\hline \\[-1.8ex] 
 & \multicolumn{3}{c}{\textit{Dependent variable:}} \\ 
\cline{2-4} 
\\[-1.8ex] & \multicolumn{3}{c}{SP500} \\ 
\\[-1.8ex] & (1) & (2) & (3)\\ 
\hline \\[-1.8ex] 
 Surprise & 0.29$^{***}$ & 0.17$^{***}$ & $-$0.09 \\ 
  & [7.74] & [2.97] & [$-$1.30] \\ 
  & & & \\ 
 Imv2 & 0.02 &  &  \\ 
  & [0.33] &  &  \\ 
  & & & \\ 
 Imv5 &  & $-$0.03 &  \\ 
  &  & [$-$0.46] &  \\ 
  & & & \\ 
 Imv10 &  &  & $-$0.07 \\ 
  &  &  & [$-$0.81] \\ 
  & & & \\ 
 CFNAI & $-$0.02 & $-$0.02 & $-$0.02 \\ 
  & [$-$0.70] & [$-$0.67] & [$-$0.65] \\ 
  & & & \\ 
 Surprise*imv2 & $-$0.23$^{***}$ &  &  \\ 
  & [$-$4.40] &  &  \\ 
  & & & \\ 
 Surprise*imv5 &  & $-$0.04 &  \\ 
  &  & [$-$0.64] &  \\ 
  & & & \\ 
 Surprise*imv10 &  &  & 0.27$^{***}$ \\ 
  &  &  & [3.29] \\ 
  & & & \\ 
 Surprise*CFNAI & $-$0.06$^{***}$ & $-$0.06$^{***}$ & $-$0.03 \\ 
  & [$-$3.09] & [$-$3.15] & [$-$1.42] \\ 
  & & & \\ 
 Constant & 0.004 & 0.04 & 0.07 \\ 
  & [0.11] & [0.70] & [0.97] \\ 
  & & & \\ 
\hline \\[-1.8ex] 
Observations & 978 & 978 & 978 \\ 
Adjusted R$^{2}$ & 0.14 & 0.12 & 0.13 \\ 
\hline 
\hline \\[-1.8ex] 
\end{tabular} 
\end{table}

% Table created by stargazer v.5.2.3 by Marek Hlavac, Social Policy Institute. E-mail: marek.hlavac at gmail.com
% Date and time: Sun, Oct 09, 2022 - 3:42:57 PM
\begin{table}[!htbp] \centering 
  \caption{\textbf{Interest rate response to MNA with an interaction term}} 
  \label{}
    \begin{flushleft}
    {\medskip\small
 The table documents the response of Treasury rates at 2, 5, and 10 years maturity to 4 major MNA surprises: $i_t = a + b Surprise_t + c MU_{t-1} + d Surprise_t MU_{t-1} + \epsilon_t.$ }
    \medskip
    \end{flushleft}
\begin{tabular}{@{\extracolsep{5pt}}lcccc} 
\\[-1.8ex]\hline 
\hline \\[-1.8ex] 
 & \multicolumn{4}{c}{\textit{Dependent variable:}} \\ 
\cline{2-5} 
\\[-1.8ex] & \multicolumn{2}{c}{$\Delta i_2$} & $\Delta i_5$ & $\Delta i_{10}$ \\ 
\\[-1.8ex] & (1) & (2) & (3) & (4)\\ 
\hline \\[-1.8ex] 
 Surprise & 2.16$^{***}$ & 0.26 & 2.12$^{***}$ & 2.16$^{***}$ \\ 
  & [17.35] & [0.74] & [5.33] & [7.53] \\ 
  & & & & \\ 
 MU &  & 1.07$^{**}$ & 1.56$^{***}$ & 0.98$^{**}$ \\ 
  &  & [2.29] & [2.95] & [2.58] \\ 
  & & & & \\ 
 Surprise*MU &  & 2.76$^{***}$ & 0.59 & $-$0.47 \\ 
  &  & [5.72] & [1.08] & [$-$1.20] \\ 
  & & & & \\ 
 Constant & $-$0.01 & $-$0.68$^{**}$ & $-$1.02$^{***}$ & $-$0.63$^{**}$ \\ 
  & [$-$0.08] & [$-$2.18] & [$-$2.88] & [$-$2.48] \\ 
  & & & & \\ 
\hline \\[-1.8ex] 
Observations & 1,008 & 1,005 & 1,005 & 1,005 \\ 
Adjusted R$^{2}$ & 0.23 & 0.26 & 0.26 & 0.26 \\ 
\hline 
\hline \\[-1.8ex] 
\textit{Note:}  & \multicolumn{4}{r}{$^{*}$p$<$0.1; $^{**}$p$<$0.05; $^{***}$p$<$0.01} \\ 
\end{tabular} 
\end{table}

%%%%%%%%%% End of results from i102PIv12 script

% Table created by stargazer v.5.2.2 by Marek Hlavac, Harvard University. E-mail: hlavac at fas.harvard.edu
% Date and time: Tue, Mar 23, 2021 - 1:36:40 PM
\begin{table}[!htbp] \centering 
  \caption{\textbf{Average returns of quintile portfolios, sorted on $\beta_{MNA}$}} 
  \label{}
  \begin{flushleft}
    {\medskip\small
 The table reports equal-weighted (ew) and value-weighted (vw) returns of quintile portfolios, formed on $\beta_{MNA}$. Each panel corresponds to a different type of MNA. I estimate $\beta_{MNA}$ using rolling-window regression of stocks returns on MNA surprises. Long-short portfolio buys stocks with the highest beta (i.e., fifth quintile) and sells stocks with the lowest beta (first quintile)}
    \medskip
    \end{flushleft}
    
\begin{tabular}{@{\extracolsep{1pt}}lcccccc}
    \toprule
    \multicolumn{7}{l}{\textbf{Panel A: Non-Farm Payroll}} \\
    \midrule  
\\[-1.8ex]\hline 
\hline \\[-1.8ex] 
 & Q1 & Q2 & Q3 & Q4 & Q5 & L/S \\ 
\hline \\[-1.8ex] 
Mean ew & $0.90^{**}$ & $0.84^{***}$ & $0.82^{***}$ & $0.83^{***}$ & $0.91^{**}$ & 0.01 \\ 
T-stat ew & [2.47] & [2.74] & [2.73] & [2.71] & [2.40] & [0.03] \\ 
Mean vw & $0.57^{*}$ & $0.77^{***}$ & $0.70^{***}$ & $0.74^{***}$ & $0.78^{**}$ & 0.21 \\ 
T-stat vw & [1.65] & [2.85] & [2.78] & [2.74] & [2.39] & [0.82] \\ 
\hline \\[-1.8ex] 
\end{tabular} 

\begin{tabular}{@{\extracolsep{1pt}}lcccccc}
    \toprule
    \multicolumn{7}{l}{\textbf{Panel B: PMI}} \\
    \midrule  
\\[-1.8ex]\hline 
\hline \\[-1.8ex] 
 & Q1 & Q2 & Q3 & Q4 & Q5 & L/S \\ 
\hline \\[-1.8ex] 
Mean ew & $0.84^{**}$ & $0.86^{***}$ & $0.96^{***}$ & $0.85^{**}$ & $0.79^{**}$ & -0.04 \\ 
T-stat ew & [2.33] & [2.92] & [3.15] & [2.55] & [2.00] & [-0.24] \\ 
Mean vw & $0.78^{**}$ & $0.80^{***}$ & $0.73^{***}$ & $0.59^{*}$ & $0.68^{**}$ & -0.10 \\ 
T-stat vw & [2.51] & [3.10] & [2.82] & [1.85] & [1.98] & [-0.51] \\ 
\hline \\[-1.8ex] 
\end{tabular} 

\begin{tabular}{@{\extracolsep{1pt}}lcccccc}
    \toprule
    \multicolumn{7}{l}{\textbf{Panel C: Retail Sales}} \\
    \midrule  
\\[-1.8ex]\hline 
\hline \\[-1.8ex] 
 & Q1 & Q2 & Q3 & Q4 & Q5 & L/S \\ 
\hline \\[-1.8ex] 
Mean ew & $0.96^{***}$ & $0.76^{**}$ & $0.89^{***}$ & $1.05^{***}$ & $1.03^{***}$ & 0.07 \\ 
T-stat ew & [2.69] & [2.48] & [2.87] & [3.25] & [2.94] & [0.43] \\ 
Mean vw & $0.86^{***}$ & $0.61^{**}$ & $0.52^{*}$ & $0.92^{***}$ & $0.98^{***}$ & 0.12 \\ 
T-stat vw & [2.75] & [2.16] & [1.80] & [3.47] & [3.31] & [0.60] \\ 
\hline \\[-1.8ex] 
\end{tabular} 

\begin{tabular}{@{\extracolsep{1pt}}lcccccc}
    \toprule
    \multicolumn{7}{l}{\textbf{Panel D: Consumer Confidence}} \\
    \midrule  
\\[-1.8ex]\hline 
\hline \\[-1.8ex] 
 & Q1 & Q2 & Q3 & Q4 & Q5 & L/S \\ 
\hline \\[-1.8ex] 
Mean ew & $0.97^{***}$ & $0.85^{***}$ & $0.83^{***}$ & $0.98^{***}$ & $1.05^{***}$ & $0.08$ \\
T-stat ew & $[3.10]$ & $[2.94]$ & $[2.85]$ & $[2.97]$ & $[2.59]$ & $[0.48]$ \\ 
Mean vw & $0.85^{***}$ & $0.66^{**}$ & $0.63^{**}$ & $0.88^{***}$ & $0.95^{***}$ & $0.10$ \\ 
T-stat vw & $[3.08]$ & $[2.51]$ & $[2.50]$ & $[2.94]$ & $[2.67]$ & $[0.53]$ \\ 
\hline \\[-1.8ex] 
\end{tabular}

\end{table}

% Table created by stargazer v.5.2.2 by Marek Hlavac, Harvard University. E-mail: hlavac at fas.harvard.edu
% Date and time: Tue, Mar 23, 2021 - 1:48:16 PM
\begin{table}[!htbp] \centering 
  \caption{\textbf{Abnormal returns of quintile portfolios, sorted on $\beta_{MNA}$}} 
  \label{} 
    \begin{flushleft}
    {\medskip\small
 The table reports abnormal returns of long-short quintile portfolios, formed on $\beta_{MNA}$. Long-short portfolio buys stocks with the highest beta (i.e., fifth quintile) and sells stocks with the lowest beta (first quintile). Each panel corresponds to a different type of MNA. MOM is the momentum factor and STR is the short-term reversal factor.}
    \medskip
    \end{flushleft}
\begin{tabular}{@{\extracolsep{-1pt}}lcccccc}
    \toprule
    \multicolumn{7}{l}{\textbf{Panel A: Non-Farm Payroll}} \\
    \midrule  
\\[-1.8ex]\hline 
\hline \\[-1.8ex] 
Statistic & Ret & $\alpha_{CAPM}$ & $\alpha_{FF3}$ & $\alpha_{Carhart}$ & $\alpha_{FF5}$ & $\alpha_{FF5+UMD+STR}$ \\ 
\hline \\[-1.8ex] 
L/S & 0.21 & 0.21 & 0.15 & 0.11 & 0.07 & 0.03 \\ 
 & [ 0.82] & [ 0.81] & [ 0.58] & [ 0.43] & [ 0.27] & [ 0.13] \\ 
\hline \\[-1.8ex]
\end{tabular} 
\begin{tabular}{@{\extracolsep{-1pt}}lcccccc}
    \toprule
    \multicolumn{7}{l}{\textbf{Panel B: PMI}} \\
    \midrule  
\\[-1.8ex]\hline 
\hline \\[-1.8ex] 
Statistic & Ret & $\alpha_{CAPM}$ & $\alpha_{FF3}$ & $\alpha_{Carhart}$ & $\alpha_{FF5}$ & $\alpha_{FF5+UMD+STR}$ \\ 
\hline \\[-1.8ex] 
L/S & -0.10 & -0.20 & -0.15 & -0.12 & -0.19 & -0.16 \\ 
 & [-0.51] & [-1.00] & [-0.79] & [-0.63] & [-0.95] & [-0.83] \\ 
\hline \\[-1.8ex] 
\end{tabular} 
\begin{tabular}{@{\extracolsep{-1pt}}lcccccc}
    \toprule
    \multicolumn{7}{l}{\textbf{Panel C: Retail Sales}} \\
    \midrule  
\\[-1.8ex]\hline 
\hline \\[-1.8ex] 
Statistic & Ret & $\alpha_{CAPM}$ & $\alpha_{FF3}$ & $\alpha_{Carhart}$ & $\alpha_{FF5}$ & $\alpha_{FF5+UMD+STR}$ \\ 
\hline \\[-1.8ex] 
L/S & 0.12 & 0.15 & 0.16 & 0.15 & 0.03 & 0.02 \\ 
& [ 0.60] & [ 0.74] & [ 0.80] & [ 0.75] & [ 0.13] & [ 0.09] \\ 
\hline \\[-1.8ex] 
\end{tabular} 
\begin{tabular}{@{\extracolsep{-1pt}}lcccccc}
    \toprule
    \multicolumn{7}{l}{\textbf{Panel D: Consumer Confidence}} \\
    \midrule  
\\[-1.8ex]\hline 
\hline \\[-1.8ex] 
Statistic & Ret & $\alpha_{CAPM}$ & $\alpha_{FF3}$ & $\alpha_{Carhart}$ & $\alpha_{FF5}$ & $\alpha_{FF5+UMD+STR}$ \\ 
\hline \\[-1.8ex] 
L/S & 0.10 & -0.10 & -0.06 & -0.06 & -0.11 & -0.10 \\ 
& [ 0.53] & [ -0.58] & [ -0.34] & [ -0.36] & [ -0.63] & [ -0.59] \\ 
\hline \\[-1.8ex]
\end{tabular} 
\end{table}

%%%%%%%%%%%%%%%%%%%%%% script PIII

% Table created by stargazer v.5.2.2 by Marek Hlavac, Harvard University. E-mail: hlavac at fas.harvard.edu
% Date and time: Tue, Mar 23, 2021 - 2:33:51 PM
\begin{table}[!htbp] \centering 
  \caption{\textbf{Announcement premium and the drift before MNA}} 
  \label{}
  \begin{flushleft}
    {\medskip\small
 The table presents average returns before and during major MNA. The first 4 columns report pre-announcement drift and announcement returns. Pre-announcement window runs from 4 pm of the previous trading day up to 10 minutes before the announcement. The announcement window is a 30-minute window around the MNA release. The last column reports a correlation between pre-announcement drift and announcement returns.}
    \medskip
    \end{flushleft}
\begin{tabular}{@{\extracolsep{5pt}} llllll} 
\\[-1.8ex]\hline 
\hline \\[-1.8ex] 
News & $Return^{Annc}$ & T-stat & $Return^{Pre-Annc}$ & T-stat & Correlation \\ 
\hline \\[-1.8ex] 
NFP & 0.03 & 0.92 & 0.10 & 3.51 & 0.05 \\ 
PMI & 0.0001 & 0.004 & -0.01 & -0.17 & 0.10 \\ 
Construction & 0.02 & 0.53 & 0.03 & 0.61 & 0.08 \\ 
Retail & 0.02 & 1.06 & -0.01 & -0.19 & -0.09 \\ 
FOMC & -0.01 & -0.29 & 0.29 & 5.25 & -0.20 \\ 
\hline \\[-1.8ex] 
\end{tabular} 
\end{table}

\pagebreak
\clearpage

\section{Appendix B: Kalman Filter} \label{sec:Model}
\setcounter{equation}{0}

\paragraph{}
This appendix describes a simple signal-extraction model, which motivates using monetary uncertainty to decompose the response of stocks to MNA shocks into a cash flow channel and monetary channel. The main result is that a change in interest rate $i_t$ is proportional to the product of monetary uncertainty $MU_{t-1}$ and growth shock $\epsilon_t$.
\paragraph{}
Consider a problem, in which the market tries to learn the value of unobserved short-term interest rate $\hat{i}_t$. To do so, it relies on its prior knowledge as well as on the signal $m_t$, contained in MNA. Thus, the problem is to dynamically use noisy signal $m_t$ to extract useful information about unobserved state $\hat{i}_t$. Suppose the law of motion of $\hat{i}_t$ and its relation to $m_t$ are as follows:
\begin{equation}
\hat{i}_t = \hat{i}_{t-1} + w_t, \hskip22em w_t \sim N(0, \sigma^2_w).
\end{equation}
\begin{equation}
m_t = \hat{i}_t + v_t, \hskip22em v_t \sim N(0, \sigma^2_m).
\end{equation}
\paragraph{}
We can solve this problem by using recursive updating and the Kalman filter. Let $\hat{i}_{t,t}$ be an estimate of unobserved rate $i_t$ for the period t, based on all observables, known at the time t. Put differently, 
\begin{equation}
\hat{i}_{t,t} = \mathbb{E}[\hat{i_t}|m_t, m_{t-1}, m_{t-2},...].
\end{equation}
Similarly, define $p_{t,t}$ as the variance of our guess of unobserved rate $\hat{i_t}$, based on all observables, known at the time t:
\begin{equation}
p_{t,t} = Var(\hat{i_t}|m_t, m_{t-1}, m_{t-2},...).
\end{equation}
The idea of the Kalman filter is to update the estimate of unknown rate $\hat{i}_{t,t}$ at every step as the weighted average of a prior estimate $\hat{i}_{t,t-1}$ and the noisy signal $m_t$. Kalman filter consists of 5 equations:
\begin{align}
K_t = \frac{p_{t,t-1}}{p_{t,t-1}+\sigma^2_{m}} \owntag {Kalman Gain}.
\end{align}
\begin{align}
\hat{i}_{t,t} = \hat{i}_{t,t-1} + K_t (m_t - \hat{i}_{t,t-1}), \owntag {State update}
\end{align}
\begin{align}
p_{t,t} = (1-K_t)p_{t,t-1}, \owntag {Variance update}.
\end{align}
\begin{align}
\hat{i}_{t+1,t} = \hat{i}_{t,t}, \owntag {State extrapolation}.
\end{align}
\begin{align}
p_{t+1,t} = p_{t,t} + \sigma^2_w, \owntag {Variance extrapolation}.
\end{align}

The state update equation implies that Kalman gain is the weight of the signal. Intuitively, Kalman gain is a ratio of prior uncertainty to prior uncertainty plus noise in a signal. Thus, when signal $m_t$ is very noisy, Kalman gain is small and we mostly use prior knowledge $p_{t,t-1}$ to estimate $\hat{i}_{t,t}$. Alternatively, a precise signal leads to high Kalman gain and we mostly rely on the signal $m_t$ to update our estimate of $i_t$. 
\paragraph{}
In general, the Kalman filter at time t works as follows:
\begin{enumerate}
    \item {Measure signal $m_t$.}
    \item {Update interest rate $\hat{i}_{t,t}$ and estimate uncertainty $p_{t,t}$ using (5), (6) and (7).}
    \item {Predict for the next period using (8) and (9).}
\end{enumerate}
\paragraph{}
We can rewrite the state update equation as

\begin{align}
\begin{split}
\hat{i}_t - \hat{i}_{t-1} & = K_t (m_t - \hat{i}_{t-1}). \\
\Delta \hat{i}_t & = \frac{p_{t,t-1}}{p_{t,t-1}+\sigma^2_{m}} (m_t - \hat{i}_{t-1}). \\
\Delta \hat{i}_t & \propto MU_{t-1} \epsilon_t.
\end{split}
\end{align}

We can think of uncertainty ${p_{t,t-1}}$ as a quantity, positively related to the uncertainty about the short-term interest rate. Then the whole Kalman Gain can be used as a proxy for monetary uncertainty $MU$. Difference between news and prior expectations $(m_t - \hat{i}_{t-1})$ is simply MNA shock $\epsilon_t$. Thus equation (B.10) implies that interest rate response is affine in an interaction between monetary uncertainty and MNA shock.

\end{document}